\documentclass[prb,twocolumn,aps,showpacs]{revtex4}
\usepackage{graphicx}
\usepackage{bm}
\usepackage{amssymb} 
\usepackage{amsmath}
\usepackage{subfigure}
\usepackage{widetext}
\usepackage{epstopdf}
\usepackage{tikz}
\usepackage{color}
\usetikzlibrary{arrows}
\newcommand{\s}{\scriptscriptstyle}

\begin{document}



\title{Magnetic resonance in slowly modulated  longitudinal field:
Modified shape of the Rabi oscillations}

\author{ R. Glenn, M. E. Limes, B. Pankovich, B. Saam,  and M. E. Raikh }
\affiliation{Department of Physics and Astronomy, University of Utah, Salt Lake City, UT 84112}

\begin{abstract}
The sensitivity of the Rabi oscillations of a resonantly driven spin-$\frac{1}{2}$ system to a weak and {\em slow} modulation of the static longitudinal magnetic field, $B_{{\scriptscriptstyle 0}}$, is studied theoretically. We establish the mapping of a weakly driven two-level system {\em with} modulation onto a {\em strongly} driven system {\em without} modulation. The mapping suggests that different regimes of spin dynamics,
known for a strongly driven system, can be realized under common experimental conditions of  weak driving (driving field $B_{\s 1}\ll B_{\s 0}$) upon proper choice of the domains of  modulation frequency, $\omega_{\s m}$, and  amplitude, $B_{\s 2}$. Fast modulation
$\omega_{\s m}\gg \Omega_{\s R}$, where $\Omega_{\s R}$ is the Rabi
frequency, emulates the regime of  driving frequency much bigger than
the resonant frequency. Strong modulation, $B_{\s 2}\gg B_{\s 1}$, emulates the regime $B_{\s 1}\gg B_{\s 0}$. Resonant modulation, $\omega_{\s m}\approx \Omega_{\s R}$, gives rise to an envelope of
the Rabi oscillations. The shape of this envelope is highly sensitive
to the detuning of the driving frequency from the resonance.
Theoretical predictions for different domains of $B_{\s 2}$ and $\omega_{\s m}$ were tested experimentally using NMR of protons in water, where, without modulation, the pattern of Rabi oscillations  could be observed over many periods. We present experimental results which reproduce the three predicted modulation regimes, and agree with theory
{\em quantitatively}.
\end{abstract}
\maketitle

\section{Introduction}

Ever since the oscillations of the
population of Zeeman levels in a
resonant ac magnetic field were predicted theoretically
by Rabi\cite{Rabi}, they have been experimentally observed in
numerous media and in various spectral ranges from the radio
to the optical. The revival of interest in  Rabi
oscillations during the past decade has been fueled by the
fact that they can now be measured in a {\em single}
two-level system (qubit).
In fact, the observation
\cite{SingleDot1,SingleDot2,SingleDot3,Josephson1,Josephson2,
Levitov0,MarcusManipulate,KoppensManipulate,
VandersypenControl,TaruchaManipulateGradient,TaruchaSingleSpinRotation,
JelezkoSingle,AwschalomCoherent,vanTolManipulate} of high-quality Rabi oscillations  in
certain isolated two-level systems such as an exciton in quantum dot\cite{SingleDot1,SingleDot2,SingleDot3},
a Josephson junction\cite{Josephson1,Josephson2,
Levitov0}, a single spin in a dot\cite{MarcusManipulate,KoppensManipulate,
VandersypenControl,TaruchaManipulateGradient,TaruchaSingleSpinRotation}, or a vacancy spin
\cite{JelezkoSingle,AwschalomCoherent,vanTolManipulate}
is viewed as  evidence that the corresponding qubit can be
coherently manipulated.

Conventional experimental conditions for observation
of the Rabi oscillations are:

\noindent{(i)} the driving frequency, $\omega$, is in resonance with the
splitting,
$\Delta_{{\scriptscriptstyle Z}}=g\mu_{{\scriptscriptstyle 0}}B_{{\scriptscriptstyle 0}}$,
of the spin levels. Here $g$ is the $g$-factor and $\mu_{{\scriptscriptstyle 0}}$ is the Bohr magneton.

\noindent{(ii)} in-plane driving field,
${\bf x}B_{{\scriptscriptstyle 1}}\cos\omega t$,
is much smaller than the static longitudinal field,
${\bf z}B_{{\scriptscriptstyle 0}}$.
This condition guarantees that the Rabi frequency,
$\Omega_{{\scriptscriptstyle R}}=g\mu_{{\scriptscriptstyle 0}}B_{{\scriptscriptstyle 1}}$,
is much smaller than $\Delta_{\s Z}$.
\begin{figure}[t]
\includegraphics[width=65mm, angle=0,clip]{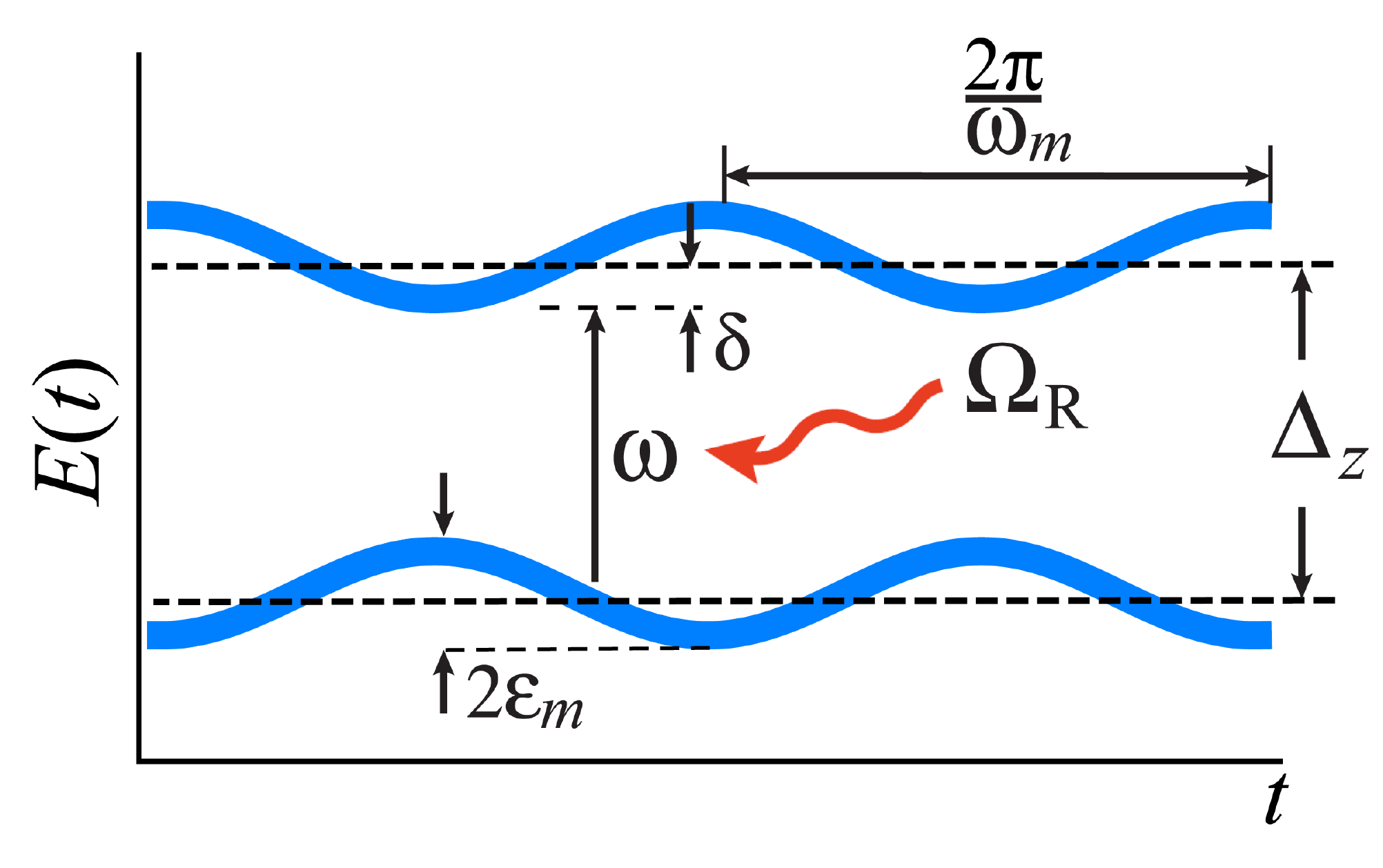}
\vspace{-0.2cm}
\caption{Schematic illustration of magnetic resonance in a modulated longitudinal field. The Zeeman splitting, $\Delta_{Z}$, oscillates in time with a small magnitude, $\varepsilon_{\s m}$, and frequency, $\omega_{\s m}$. }
\label{fig:intro}
\end{figure}
Under the conditions (i), (ii)
the classical Rabi result\cite{Rabi}
\begin{equation}
\label{Rabi}
P_{+\frac{1}{2}}(t)=\frac{\Omega_{{\scriptscriptstyle R}}^2}{\omega_{{\scriptscriptstyle 0}}^2}\sin^2\frac{\omega_{{\scriptscriptstyle 0}}t}{2},
\end{equation}
for the oscillating population
of the upper Zeeman level, which
was empty at $t=0$, applies.
Here
\begin{equation}
\label{frequency}
\omega_{{\scriptscriptstyle 0}}=\sqrt{\delta^2+\Omega_{{\scriptscriptstyle R}}^2}
\end{equation}
is the oscillation frequency, and
$\delta=\Delta_{{\scriptscriptstyle Z}}-\omega$ is a detuning
of the driving frequency from  resonance.

It is known\cite{Townes,Shirley} that almost
sinusoidal Rabi oscillations can also be excited
in the domains of parameters where the conditions (i)
and (ii) are not met. One example\cite{Townes} is the
multiphoton Rabi oscillations which develop when
the driving frequency is close to
\begin{equation}
\label{multiphoton}
\omega^{(p)}=\frac{\Delta_{\s Z}}{2p+1},
\end{equation}
where $p$ is positive integer.  Another
example\cite{Shirley} is realized in the domain of frequencies
\begin{equation}
\label{domain}
\omega \sim \Omega_{\s R}\gg \Delta_{\s Z}.
\end{equation}
Existence of the domain Eq. (\ref{domain})
requires a strong driving field $B_{\s 1}\gg B_{\s 0}$,
which is quite unusual for magnetic resonance.

The main point of the present paper is that the entire spectrum of
regimes of the Rabi oscillations can be realized
when conditions (i) and (ii) still apply, but the longitudinal
field contains a small {\em slowly} oscillating component,
${\bf z}B_{{\scriptscriptstyle 2}}(t)$, where $B_{\s 2}\ll B_{\s 0}$.
In other words, we will assume that the Zeeman splitting, $\Delta_{\s Z}$, has a time dependent correction,
\begin{equation}
\varepsilon(t)=g\mu_{{\scriptscriptstyle 0}}B_{{\scriptscriptstyle
2}}(t),
\end{equation}
and study how the classical result Eq. (\ref{Rabi}) is affected
by this correction. We will demonstrate that even when the magnitude of
the correction is much smaller than $\Delta_{\s Z}$ and it oscillates
with a frequency much smaller than $\Delta_{\s Z}$, see Fig. \ref{fig:intro}, its effect on the
Rabi oscillations can still be dramatic. To prove this statement in Sect. II
we map a weakly and resonantly driven two-level system {\em with} modulation onto a {\em strongly} driven system {\em without} modulation. In sections III, IV, and V we make use of this mapping to study how
different regimes of spin dynamics emerge in different limits of
modulation, namely, fast (compared to $\Omega_{\s R}$) modulation,
strong (compared to $\Omega_{\s R}$) modulation, and near-resonant modulation (with frequency $\approx \Omega_{\s R}$).
In Sect. VI we report the results of the experimental
test of theoretical predictions.
NMR measurements on protons in water have advantage that, without modulation,
the pattern of Rabi oscillations  could be observed over many periods.
By applying the modulation pulse with varying frequency and magnitude,
we were able to reproduce the above three regimes of spin dynamics.
Concluding remarks are presented in Sect. VII.

\section{Mapping onto a strongly driven system without modulation}

The time evolution of the amplitudes $C_{+\frac{1}{2}}$ and $C_{-\frac{1}{2}}$ of two spin orientations
is governed by the system of equations
\begin{equation}
\label{basic1}
i{\dot{C}}_{+\frac{1}{2}}=\frac{\Delta_{{\scriptscriptstyle Z}}+\varepsilon(t)}{2}C_{+\frac{1}{2}}+
\Omega_{{\scriptscriptstyle R}}C_{-\frac{1}{2}}\cos\omega t ,
\end{equation}
\begin{equation}
\label{basic2}
i{\dot{C}}_{-\frac{1}{2}}=-\frac{\Delta_{{\scriptscriptstyle Z}}+\varepsilon(t)}{2}C_{-\frac{1}{2}}+
\Omega_{{\scriptscriptstyle R}}C_{+\frac{1}{2}}\cos\omega t .
\end{equation}
In the absence of modulation, $\varepsilon (t)=0$, it is convenient to reduce
this system to a single second-order differential equation by
introducing, instead of $C_{+\frac{1}{2}}$ and $C_{-\frac{1}{2}}$,
the combinations
\begin{equation}
\label{A+-}
A_+=C_{+\frac{1}{2}}+C_{-\frac{1}{2}}, ~~~~~A_-=C_{+\frac{1}{2}}-C_{-\frac{1}{2}}.
\end{equation}
The system of equations for the new amplitudes reads
\begin{equation}
\label{A+dot}
i{\dot{A}_+}=\frac{\Delta_{{\scriptscriptstyle Z}}}{2}A_- +
\Omega_{{\scriptscriptstyle R}}A_+\cos\omega t,
\end{equation}

\begin{equation}
\label{A-dot}
i{\dot{A}_-}=\frac{\Delta_{{\scriptscriptstyle Z}}}{2}A_+ -
\Omega_{{\scriptscriptstyle R}}A_-\cos\omega t.
\end{equation}
Upon expressing $A_-$ from Eq. (\ref{A+dot}) and substituting it into Eq. (\ref{A-dot}),
we arrive at the sought equation
\begin{equation}
\label{A+dotdot}
{\ddot{A}_+}+\Bigl[-i\omega\Omega_{{\scriptscriptstyle R}}\sin\omega t +\Omega_{{\scriptscriptstyle R}}^2\cos^2\omega t +\frac{\Delta_{{\scriptscriptstyle Z}}^2}{4}\Bigr]A_+\!=0.
\end{equation}

 Assume now that the modulation is present, but, due to the weakness of the ac field,
 $\Omega{{\scriptscriptstyle R}}\ll \Delta_{{\scriptscriptstyle Z}}$, the system
 Eqs. (\ref{basic1}), (\ref{basic2}) can be treated within the rotating-wave approximation (RWA).
 This amounts to the replacement of $\cos\omega t$ by $\frac{1}{2}e^{-i\omega t}$ in Eq. (\ref{basic1}),
and by $\frac{1}{2}e^{i\omega t}$ in Eq. (\ref{basic2}).
Then, upon standard transformation into the rotating system
\begin{equation}
\label{rotating}
C_{+\frac{1}{2}}(t)=D_{+\frac{1}{2}}(t)e^{-\frac{i}{2}\omega t},~~~C_{-\frac{1}{2}}(t)=D_{-\frac{1}{2}}(t)e^{\frac{i}{2}\omega t},
\end{equation}
these equations take the form
\begin{equation}
\label{D1}
i{\dot{D}}_{+\frac{1}{2}}=\frac{\delta+\varepsilon(t)}{2}D_{+\frac{1}{2}}+
\frac{\Omega_{{\scriptscriptstyle R}}}{2}D_{-\frac{1}{2}},
\end{equation}

\begin{equation}
\label{D2}
i{\dot{D}}_{-\frac{1}{2}}=-\frac{\delta+\varepsilon(t)}{2}D_{-\frac{1}{2}}+
\frac{\Omega_{{\scriptscriptstyle R}}}{2}D_{+\frac{1}{2}}.
\end{equation}
As a next step, we express $D_{-\frac{1}{2}}$ from Eq. (\ref{D1}) and substitute it into Eq. (\ref{D2}).
This leads to the following second-order differential equation
for $D_{+\frac{1}{2}}$

\begin{equation}
\label{secondorder}
{\ddot{D}}_{+\frac{1}{2}}+\Bigl[i\frac{\dot{\varepsilon}(t)}{2}
+\frac{(\delta+\varepsilon)^2+
\Omega_{{\scriptscriptstyle R}}^2}{4}\Bigr]D_{+\frac{1}{2}}=0.
\end{equation}
At this point we assume that  the modulation is sinusoidal,
\begin{equation}
\label{modulation}
\varepsilon(t)=\varepsilon_{{\scriptscriptstyle m}}\cos\omega_{{\scriptscriptstyle m}}t,
\end{equation}
and make the key observation that, for $\delta=0$,  Eq. (\ref{secondorder}) reduces to Eq. (\ref{A+dotdot})
upon replacement
\begin{equation}
\label{correspondence}
\Omega_{{\scriptscriptstyle R}} \rightarrow \Delta_{{\scriptscriptstyle Z}},~~\varepsilon_{{\scriptscriptstyle m}}\rightarrow 2\Omega_{{\scriptscriptstyle R}},~~
\omega_{{\scriptscriptstyle m}} \rightarrow \omega.
\end{equation}
This mapping provides an important insight into the effect of modulation of the longitudinal
field on magnetic resonance. Indeed, Eq. (\ref{A+dotdot}) captures the time evolution of the
populations $|C_{+\frac{1}{2}}(t)|^2$ and $|C_{-\frac{1}{2}}(t)|^2$
when
the ac drive is not weak, i.e.,
$\Omega_{{\scriptscriptstyle R}} \gtrsim \Delta_{{\scriptscriptstyle Z}}$, so that RWA does not apply.
On the other hand, Eq. (\ref{secondorder}) describes the spin dynamics with {\em weak}
ac drive, but in the presence of the modulation.
Solutions of Eq.  (\ref{A+dotdot}), the structure of which is dictated by the Floquet theorem, were analyzed
in a great number of papers starting from pioneering works\cite{Townes,Shirley}.
Therefore, the approaches developed for a  driven
spin-$\frac{1}{2}$ system without modulation can be
utilized for the case when the modulation is present.

For example, in the absence of modulation, resonant drive
of the spin-$\frac{1}{2}$ system corresponds to the condition
$\omega=\Delta_{\s Z}$. Then from Eq. (\ref{correspondence})
we conclude that the effect of modulation on the Rabi oscillations is most pronounced when the modulation frequency,  $\omega_{{\scriptscriptstyle m}}$, is close to the Rabi frequency, $\Omega_{{\scriptscriptstyle R}}$.
This sensitivity was previously pointed out in
Refs. \onlinecite{rotary,DoublyDriven,Greenberg1,Greenberg2,Saiko1}.

A less obvious consequence of the mapping Eq. (\ref{correspondence})
is that, upon increasing the modulation strength,
$\varepsilon_{{\scriptscriptstyle m}}$,
the Rabi oscillations become sensitive to the modulation
for  modulation frequencies
 close to
\begin{equation}
\label{fractional}
\omega^{(p)}_{{\scriptscriptstyle m}}= \frac{\Omega_{{\scriptscriptstyle R}}}{2p+1}.
\end{equation}
Condition Eq. (\ref{fractional}) implies that
the modulation affects the Rabi oscillations when
the modulation period contains an odd number
of the Rabi periods. This condition
is an analog of the condition for {\em multiphoton}
resonances in a driven two-level system\cite{Townes}
which take place at ac driving frequencies
$\omega \approx \omega^{(p)}$, where $\omega^{(p)}$ is given by Eq.~(\ref{multiphoton}).

Despite the fact that Eqs. (\ref{A+dotdot}) and (\ref{secondorder})
can be formally mapped onto each other, the physical phenomena that
they describe are vastly different.
For the most prominent example, $\omega\approx\Delta_{\s Z}$,
Eq. (\ref{A+dotdot}) describes the Rabi oscillations of the
populations of the Zeeman levels, while
Eq. (\ref{secondorder}) describes  {\em slow envelope} of these
oscillations, which emerges due to modulation, $\varepsilon(t)$,
of $\Delta_{{\scriptscriptstyle Z}}$.
Formally, the difference
stems from the initial conditions, which must be imposed on the solutions
of Eqs. (\ref{A+dotdot}) and (\ref{secondorder}).
If, for example, at $t=0$ the spin points down, then the initial
conditions for Eq. (\ref{A+dotdot}) read
\begin{equation}
\label{initialA}
A_+(0)=1,~~~{\dot{A}_+}(0)=i\left(\frac{\Delta_{{\scriptscriptstyle Z}}}{2}-\Omega_{{\scriptscriptstyle R}}   \right).
\end{equation}
For the same initial state, the initial conditions for Eq. (\ref{secondorder}) have the form
\begin{equation}
\label{initialD}
D_{+\frac{1}{2}}(0)=0,~~~{\dot{D}}_{+\frac{1}{2}}(0)=-i\frac{\Omega_{{\scriptscriptstyle R}}}{2}.
\end{equation}
Below we study  the spin dynamics, $P_{+\frac{1}{2}}(t)$,  in different domains
of modulation frequencies and magnitudes.

\section{ Fast modulation: $\omega_{\s m}\gg  \Omega_{\s R}$.}
\label{Fast Modulation}
Qualitatively, it is clear that weak modulation $\varepsilon_{\s m}\ll \omega_{\s m}$
averages out if the modulation frequency is much bigger than $\Omega_{\s R}$. Then the Rabi oscillations remain unaffected. Nontrivial modification of the Rabi
 oscillations takes place when both  $\omega_{\s m}$ and $\varepsilon_{\s m}$
 are much bigger than $\Omega_{\s R}$. For conventional magnetic resonance this
 regime corresponds to
 $\omega \gg \Delta_{\s Z}$ and $\Omega_{\s R}\gg \Delta_{\s Z}$, see Eq.
 (\ref{correspondence}), i.e., $B_{\s 1}\gg B_{\s 0}$, which is exotic. Fast modulation,
 on the other hand, is fully compatible with $B_{\s 1}\ll B_{\s 0}$.

For simplicity we will consider the case of zero detuning, $\delta=0$. To analyze the limit $\omega_{{\scriptscriptstyle m}}\gg  \Omega_{{\scriptscriptstyle R}}$
it is  convenient to introduce the new variables
\begin{align}
\label{newvariables}
&{\tilde D}_{-\frac{1}{2}}\!=\!D_{-\frac{1}{2}}(t)e^{\frac{i\varepsilon_{\s m}}{2\omega_{\s m}}\sin\omega_{\s m}t},&\nonumber\\
&{\tilde D}_{+\frac{1}{2}}\!=\!D_{+\frac{1}{2}}(t)e^{\frac{-i\varepsilon_{\s m}}{2\omega_{\s m}}\sin\omega_{\s m}t},&
\end{align}
and to rewrite the system Eqs. (\ref{D1}), (\ref{D2}) as
\begin{align}
\label{dottildeD1}
&i{\dot{\tilde D}}_{-\frac{1}{2}}=\frac{\Omega_{\s R}}{2}
e^{\frac{i\varepsilon_{\s m}}{\omega_{\s m}}\sin\omega_{\s m}t}{\tilde D}_{+\frac{1}{2}},&\nonumber\\
&i{\dot{\tilde D}}_{+\frac{1}{2}}=\frac{\Omega_{\s R}}{2}
e^{-\frac{i\varepsilon_{\s m}}{\omega_{\s m}}\sin\omega_{\s m}t}{\tilde D}_{-\frac{1}{2}}.&
\end{align}
\begin{figure}[t]
\includegraphics[width=77mm,clip]{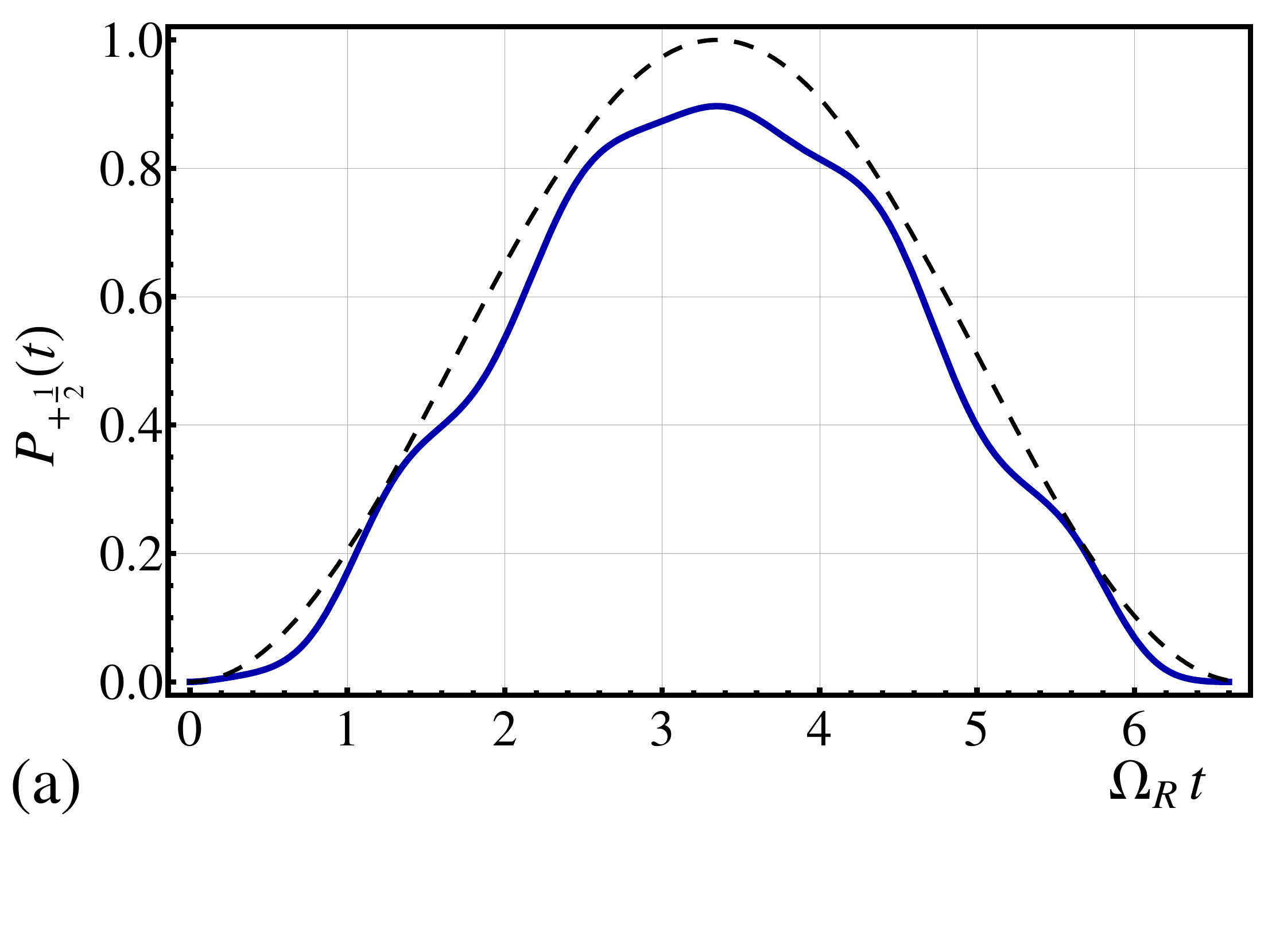}
\vspace{-0.8cm}
\includegraphics[width=77mm,clip]{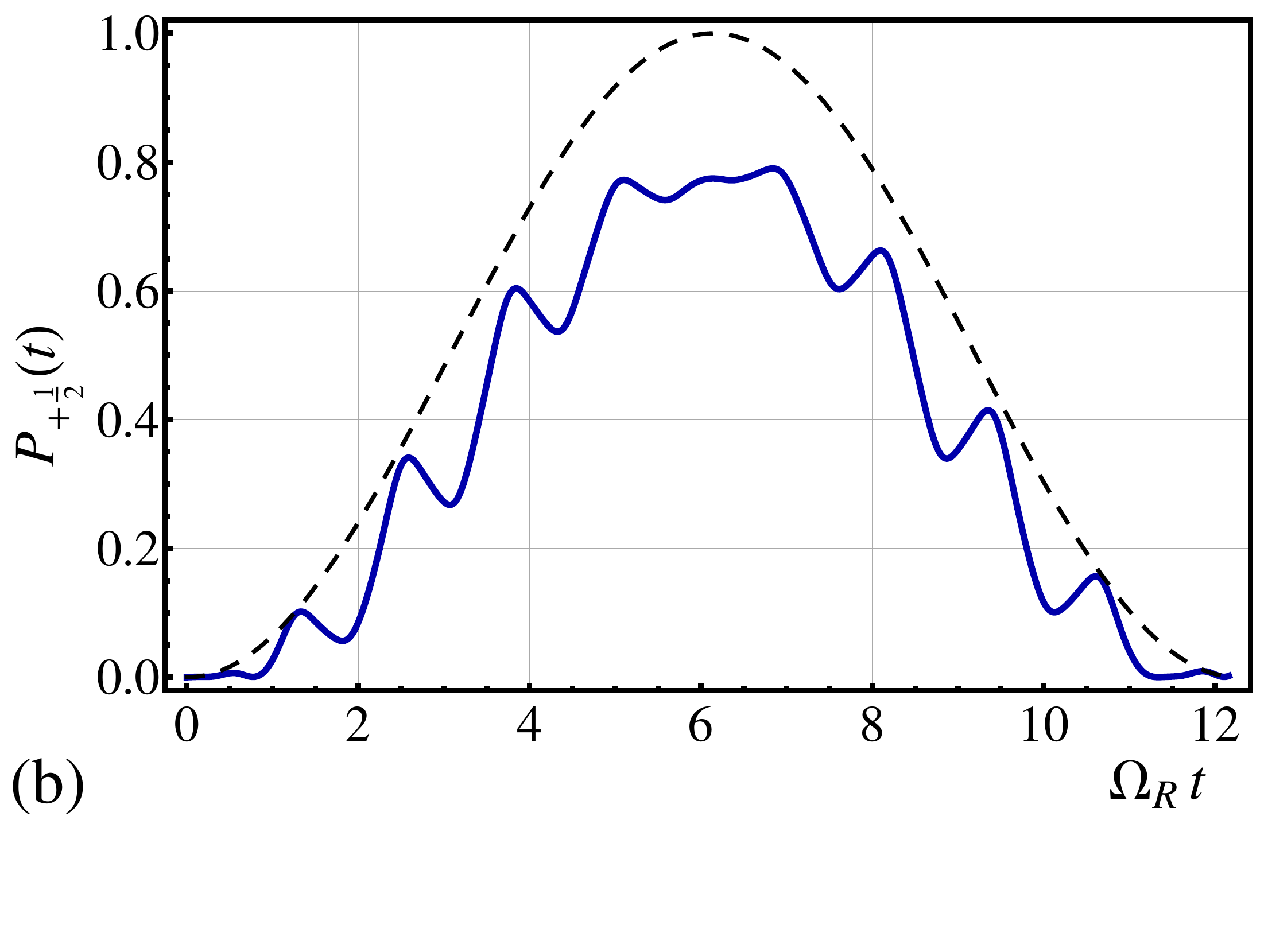}
\vspace{-0.8cm}
\includegraphics[width=77mm,clip]{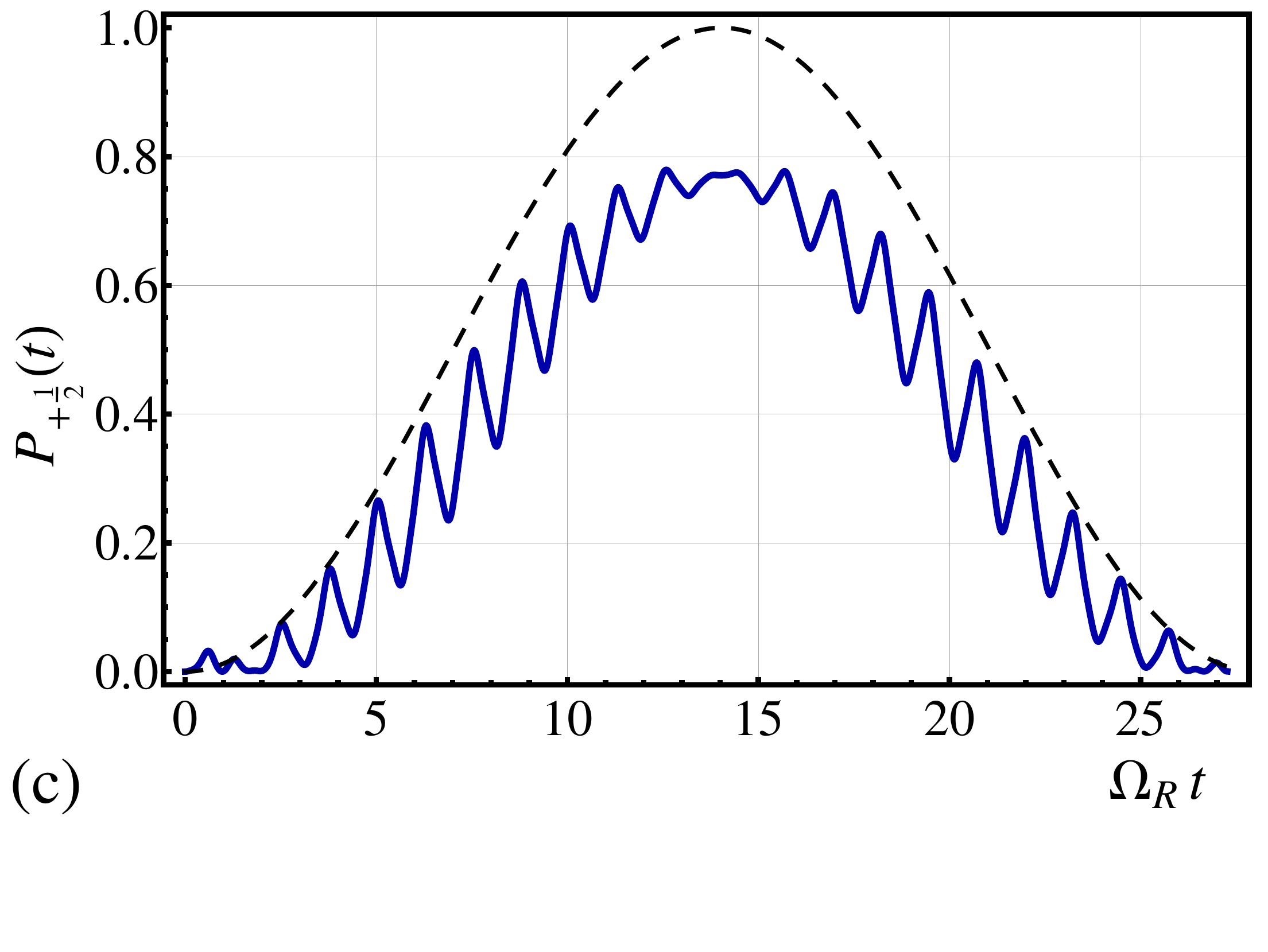}
\vspace{-0.8cm}
\caption{Fast modulation regime,
$\omega_{\s m}\gg \Omega_{\s R}$.
Modulation causes slowing down of the Rabi
oscillations, which are plotted with dashed
lines from  Eq. \eqref{Modiefied} for
$\omega_{\s m}=5\Omega_{\s R}$
and three values of the ratio  $\frac{\varepsilon_{\s m}}{\omega_{\s m}}$: (a) $0.5$, (b) $1.5$, (c) $2$. Correction Eq. \eqref{correction} gives rise to the fast component of the Rabi oscillations. The
probability, $P_{+\frac{1}{2}}(t)$, with fast
component taken into account is shown with full lines.}
\label{fig:fastmod}
\end{figure}
\!\!If we now replace $\exp{\bigl(\pm \frac{i\varepsilon_{\s m}}{\omega_{\s m}}\sin\omega_{\s m}t}\bigr)$ by the time average, $J_{\s 0}\!\!\left(\!\frac{\varepsilon_{\s m}}{\omega_{\s m}}\!\right)$, where $J_{\s 0}(z)$ is a zero-order Bessel function, the system Eq. (\ref{dottildeD1}) will readily yield
\begin{equation}
\label{Solution}
{\tilde D}_{-\frac{1}{2}}\!=\!\cos\left(\frac{\Omega_{\s R}}{2}J_{\s 0}\!\!\left(\!\frac{\varepsilon_{\s m}}{\omega_{\s m}}\!\right)t\right)\!,~
{\tilde D}_{+\frac{1}{2}}\!=-i\!\sin\left(\frac{\Omega_{\s R}}{2}J_{\s 0}\!\!\left(\!\frac{\varepsilon_{\s m}}{\omega_{\s m}}\!\right)t\right)\!,
\end{equation}
so that
\begin{equation}
\label{Modiefied}
P_{+\frac{1}{2}}(t)=\sin^2\left(\frac{\Omega_{\s R}}{2}J_{\s 0}\!\!\left(\!\frac{\varepsilon_{\s m}}{\omega_{\s m}}\!\right)t\right).
\end{equation}
The latter expression suggests that the effect of fast modulation,
$\omega_{\s m}\gg \Omega_{\s R}$, on the Rabi oscillations
is the reduction of their frequency by a factor
$J_{\s 0}\!\!\left(\!\frac{\varepsilon_{\s m}}{\omega_{\s m}}\!\right)$.
The reduction is significant when the modulation amplitude, $\varepsilon_{\s m}$, is of the order of $\omega_{\s m}$.

The remaining task is to demonstrate that, for arbitrary modulation strength, $\varepsilon_{\s m}$,  the condition $\omega_{\s m} \gg \Omega_{\s R}$ justifies the replacement of $\exp{\bigl(\pm \frac{i\varepsilon_{\s m}}{\omega_{\s m}}\sin\omega_{\s m}t}\bigr)$
by $J_{\s 0}\!\!\left(\!\frac{\varepsilon_{\s m}}{\omega_{\s m}}\!\right)$.
For this purpose we consider the correction
to ${\tilde D}_{+\frac{1}{2}}$ coming from higher harmonics
of $\exp{\bigl(\frac{i\varepsilon_{\s m}}{\omega_{\s m}}\sin\omega_{\s m}t}\bigr)$. It is convenient to cast this correction in the form

\begin{align}
\label{correction}
&\hspace{-0.1cm}{\tilde D}_{+\frac{1}{2}}+i\sin\left(\frac{\Omega_{\s R}}{2}J_{\s 0}\!\!\left(\!\frac{\varepsilon_{\s m}}{\omega_{\s m}}\!\right)t\right)&
\nonumber\\
&\hspace{-0.1cm}=-\frac{i\Omega_{\s R}}{2\omega_{\s m}}\int_0^{\omega_{\s m}t}
\!\!d\phi\left[e^{-\frac{i\varepsilon_{\s m}}{\omega_{\s m}}\sin\phi}-
\!\!J_{\s 0}\!\!\left(\!\frac{\varepsilon_{\s m}}{\omega_{\s m}}\!\right)\right]\cos\eta\phi,&
\end{align}
where we introduced a small parameter
\begin{equation}
\label{gamma}
\eta = \frac{\Omega_{\s R}}{2\omega_{\s m}}J_{\s 0}\!\!\left(\!\frac{\varepsilon_{\s m}}{\omega_{\s m}}\!\right) \ll 1.
\end{equation}
There is a small prefactor in front of the integral.
Still one has to check that the integral does not grow at large $\omega_{\s m}t$. Suppose that $\omega_{\s m}t=2\pi N$.
Then the integration over $N$ modulation periods can be reduced to a single integral from $0$ to $2\pi$ in which $\cos\eta\phi$ is replaced by

\begin{align}
\label{sum}
&\sum_{n=0}^N\!\cos\eta(\phi-2\pi n)=\frac{\cos\eta\phi}{2}\left(\!\frac{\sin2\pi \eta N}{\tan\pi\eta}\!+\!\cos2\pi\eta N+1\!\right)&
\nonumber\\
&\hspace{0.25cm}+\frac{\sin\eta\phi}{2}\left(\frac{1-\cos2\pi\eta N}{\tan\pi\eta}+\sin2\pi\eta N\right)\!.&
\end{align}
We see that for $N<\eta^{-1}$ the first term grows with $N$. Moreover, this term is much bigger than the second term. However, the correction Eq. (\ref{correction}) is determined by the
second term in Eq. (\ref{sum}). The reason is that the first term
does not depend on $\phi$ with accuracy $\eta$. On the other hand,
the part of the first term which is independent of $\phi$ {\em does not} contribute to the correction Eq. (\ref{correction}). This is because the integral from this part is zero.
The second term in Eq. (\ref{sum}), by virtue of smallness of  $\eta$, can be replaced by $\frac{\phi}{\pi}\sin^2\pi\eta N$, so it remains finite at large $N$. This proves that, even for strong modulation, the correction Eq. (\ref{correction}) is $\sim \Omega_{\s R}/\omega_{\s m}$.
In terms of the probability $P_{+\frac{1}{2}}(t)$,
the correction Eq. (\ref{correction}) gives rise to
a fast component, as illustrated in Fig.~\ref{fig:fastmod}.

\section{Strong modulation: $ \varepsilon_{\s m}\gg\Omega_{\s R}\gg \omega_{\s m}$}
\label{Strong Modulation}
For strong modulation, $\varepsilon_{\s m}\gg \Omega_{\s R}$, the
term $\Omega_{\s R}^2$ in Eq. (\ref{secondorder}) is  small compared
to $\varepsilon^2$, which suggests that Rabi oscillations do not
develop. However, if the modulation is slow enough, and the
detuning, $\delta$, is
much smaller than $\Omega_{\s R}$, the term $\Omega_{\s R}^2$ will be
dominant during short (compared to $\omega_{\s m}^{-1}$) time intervals
near
\begin{equation}
\label{tk}
 t=t_{\s k}=\left(\frac{\pi}{2}+\pi k\right)\omega_{\s m}^{-1},
\end{equation}
when $\varepsilon(t)$ passes through zero. Within these intervals we can set
$t=t_{\s k}+t_1$ and expand $\varepsilon(t)$ with respect to $t_1$. Then
Eq. (\ref{secondorder}) assumes the form
\begin{equation}
\label{secondorder1}
{\ddot{D}}_{+\frac{1}{2}}+\Biggl[i\frac{\varepsilon_{\s m}\omega_{\s m}}{2}
+\frac{\varepsilon_{\s m}^2\omega_{\s m}^2t_{\s 1}^2+
\Omega_{{\scriptscriptstyle R}}^2}{4}\Biggr]D_{+\frac{1}{2}}=0,
\end{equation}
where we assumed for definitiveness that $k$ is odd.

Solution of Eq. (\ref{secondorder1}) can be expressed in terms of parabolic
cylinder functions\cite{Transcendental}
\begin{equation}
\label{parabolic}
D_{+\frac{1}{2}}(t_{\s 1})=f_{\s 1}\,{\cal D}_{\nu}\!\left(\!e^{\frac{\pi i}{4}}\frac{t_{\s 1}}{\tau}\!\right)+ f_{\s 2}\,{\cal D}_{\nu}\!\left(\!\!-e^{\frac{\pi i}{4}}\frac{t_{\s 1}}{\tau}\!\right)\!,
\end{equation}
where the characteristic time, $\tau$, is given by
\begin{equation}
\tau=\frac{1}{(\varepsilon_{\s m}\,\omega_{\s m})^{1/2}},
\end{equation}
and parameter $\nu$ is defined as
\begin{equation}
\label{nu}
\nu=-i~\frac{\Omega_{\s R}^2}{4\varepsilon_{\s m}\,{\omega_{\s m}}}.
\end{equation}
Expansion of $\varepsilon(t)$ is justified under the condition,
$\omega_{\s m}\tau \ll 1$, which reduces to $\varepsilon_{\s m}\gg \omega_{\s m}$.
With regard to parameter, $\nu$, it can be presented as a product
$|\nu|=\left(\frac{\Omega_{\s R}}{2\varepsilon_{\s m}}\right)^2\!\!\left(\frac{\varepsilon_{\s m}}{\omega_{\s m}}\right)$.
The first factor of this product is small, while the second factor is
large. This means that in the domain
$\varepsilon_{\s m}\gg\Omega_{\s R}\gg \omega_{\s m}$ the value $|\nu|$ can be both large and small. The magnitude of $|\nu|$ determines the angle of the spin
rotation as $\varepsilon(t)$ passes through zero. To see this, assume that at
moment $t=t_{\s k}$ the spin points down.
Then the constants $f_{\s 1}$, $f_{\s 2}$ in Eq. (\ref{parabolic})
can be found from
the initial conditions Eq. (\ref {initialD}), yielding
\begin{equation}
\label{final}
D_{+\frac{1}{2}}(t_{\s 1})\!=\frac{|\nu|^{1/2}\,\, e^{-\frac{3\pi i}{4}}}{2\,{\cal D}_{\nu}^{\prime}(0)}
\left[
{\cal D}_{\nu}\!\left(\!e^{\frac{\pi i}{4}}\frac{t_{\s 1}}{\tau}\right)
-
{\cal D}_{\nu}\!\left(\!\!-e^{\frac{\pi i}{4}}\frac{t_{\s 1}}{\tau}\right)
\right]\!,
\end{equation}
where ${\cal D}_{\nu}^{\prime}(0)$ is the derivative of ${\cal D}_{\nu}(z)$ at
$z=0$.
The dependence of ${\cal D}_{\nu}^{\prime}(0)$ on $\nu$ can be
established from the integral representation of the parabolic
cylinder function\cite{Transcendental}
\begin{equation}
\label{derivative}
{\cal D}_{\nu}^{\prime}(0)=\biggl(\frac{2}{\pi}\biggr)^{1/2}2\,^{\nu/2}\sin\left(\frac{\pi \nu}{2}\right) \Gamma\left(\frac{\nu}{2}+1\right)\!.
\end{equation}
Here $\Gamma(z)$ is the gamma-function. The probability $P_{+\frac{1}{2}}(t_{\s 1})$
contains $|{\cal D}_{\nu}^{\prime}(0)|^2$, which, using the properties of the gamma-function, can be simplified to

\begin{equation}
\label{absolute}
|{\cal D}_{\nu}^{\prime}(0)|^2=|\nu|\sinh\!\left(\!\frac{\pi|\nu|}{2}\!\right).
\end{equation}
Substituting Eq. (\ref{absolute}) into Eq. (\ref{final}), we arrive at the final
result
\begin{equation}
\label{probability}
P_{+\frac{1}{2}}(t_{\s 1})=\frac{1}{4\sinh\!\left(\!\frac{\pi|\nu|}{2}\!\right)}
\left\vert
{\cal D}_{\nu}\!\left(\!e^{\frac{\pi i}{4}}\frac{t_{\s 1}}{\tau}\!\right)
-
{\cal D}_{\nu}\!\left(\!\!-e^{\frac{\pi i}{4}}\frac{t_{\s 1}}{\tau}\!\right)
\right\vert^2.
\end{equation}
\begin{figure}[t]
\includegraphics[width=77mm,clip]{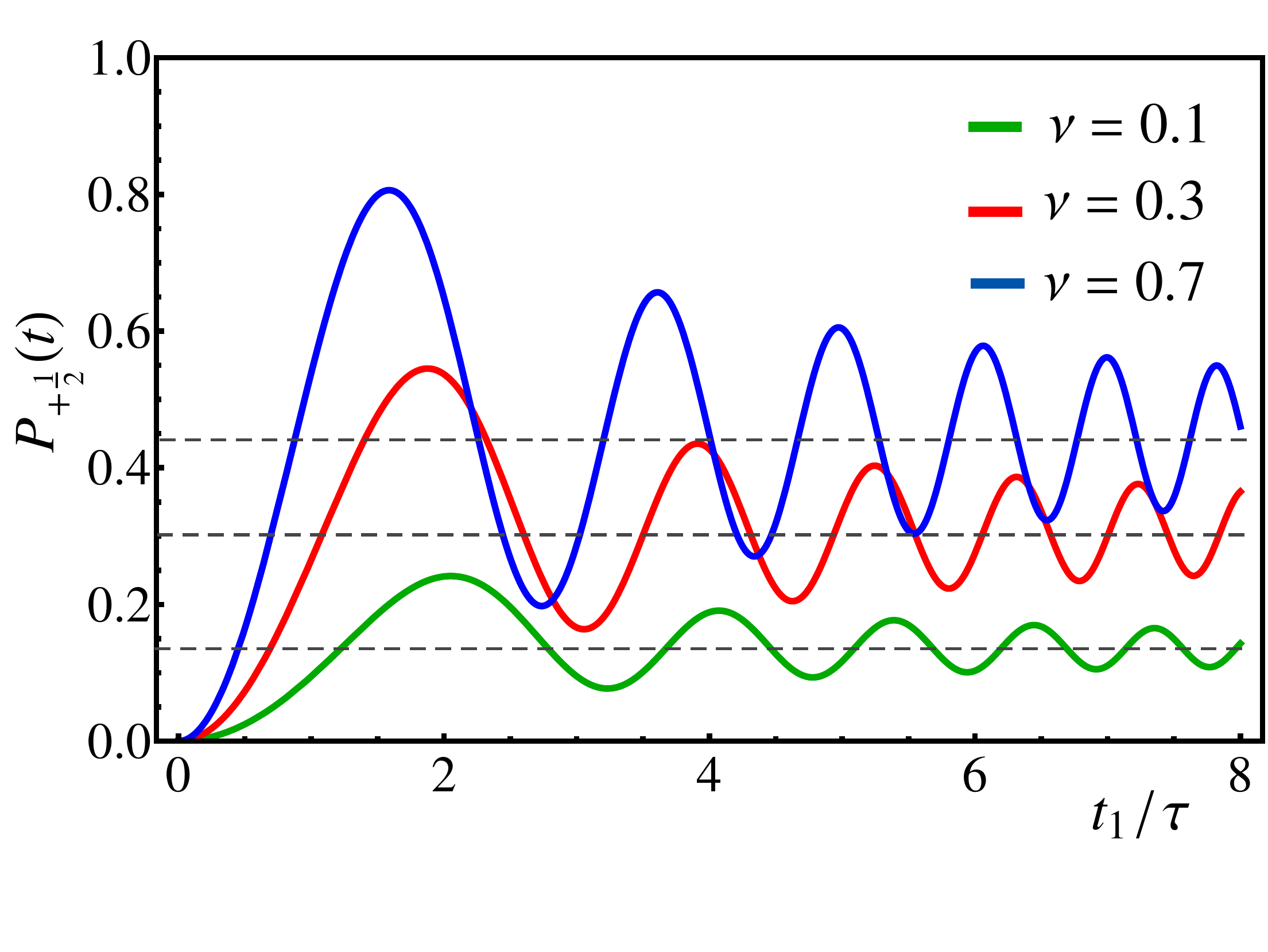}
\vspace{-0.8cm}
\caption{Strong modulation regime,
$\varepsilon_{\s m}\gg\Omega_{\s R}\gg \omega_{\s m}$.
Spin dynamics persists only during short intervals
$|t-t_{\s k}|\sim \tau \ll \omega_{\s m}^{(-1)}$
around the moments, $t_{\s k}$, when modulation passes through zero.
 Probability, $P_{+\frac{1}{2}}$, is plotted from Eq. \eqref{probability} versus dimensionless
 time $(t-t_{\s k})/\tau$ for three values
 of parameter  $|\nu|$, Eq. (\ref{nu}),  describing the strength of the ac drive.
 Green, red, and blue curves correspond to the
 values  $|\nu|=0.1$, $|\nu|=0.3$, and $|\nu|=0.7$, respectively.}
 \label{fig:strongmod}
\end{figure}

In Fig.~\ref{fig:strongmod} the dependence $P_{+\frac{1}{2}}(t_{\s 1})$ is plotted
from Eq. (\ref{probability}) for different values of the parameter $|\nu|$.
We see that, starting from zero, $P_{+\frac{1}{2}}(t_{\s 1})$ oscillates with
$t_{\s 1}$, and, finally, saturates at $t_{\s 1}\gg \tau$ at some finite value, $P_{+\frac{1}{2}}(\infty)$.
To find $P_{+\frac{1}{2}}(\infty)$ we use the large-$z$ asymptote
of the parabolic cylinder function\cite{Transcendental}, ${\cal D}_{\nu}(z) \sim z^{\nu}\exp{\left(\frac{1}{4}z^2\right)}$, and the fact that
the asymptotes of ${\cal D}_{\nu}(z)$ and
${\cal D}_{\nu}(-z)$ differ by $\exp(i\pi\nu)=\exp(-\pi|\nu|)$.
This allows to find the saturation level of the numerator to be $4\sinh^2\!\left(\frac{\pi|\nu|}{4}\right)\exp\left(\!-\frac{\pi|\nu|}{2}\right)$.
Subsequently, the saturation level of $P_{+\frac{1}{2}}$ assumes the form
\begin{equation}
\label{saturation}
P_{+\frac{1}{2}}(\infty)=\frac{1-e^{-\pi|\nu|}}{2}.
\end{equation}
The meaning of the result Eq. (\ref{saturation}) is transparent.
If the driving field is weak, so that $|\nu|\ll 1$, we have $P_{+\frac{1}{2}}(\infty)\ll 1$, which implies that
the spin pointing down at
$t=t_{\s k}$ retains its orientation throughout the entire interval
$t_{\s k+1}-t_{\s k}=\frac{\pi}{\omega_{\s m}}$. If the driving field
is strong, so that $|\nu|\gg 1$, the spin, during the short time $\ll \frac{\pi}{\omega_{\s m}}$,  will rotate to the position
$P_{+\frac{1}{2}} \approx P_{-\frac{1}{2}} \approx \frac{1}{2}$, i.e.,
will point along the $x$-axis,  and spend in this position the remaining
part of the interval  $t_{\s k+1}-t_{\s k}$.

\section {Weak resonant modulation: $\omega_{{\scriptscriptstyle m}}\approx  \Omega_{{\scriptscriptstyle R}}; \varepsilon_{{\scriptscriptstyle m}}\ll \Omega_{{\scriptscriptstyle R}}.$}
\label{Weak-resonant Modulation}
\begin{figure}[t]
\includegraphics[width=77mm, angle=0,clip]{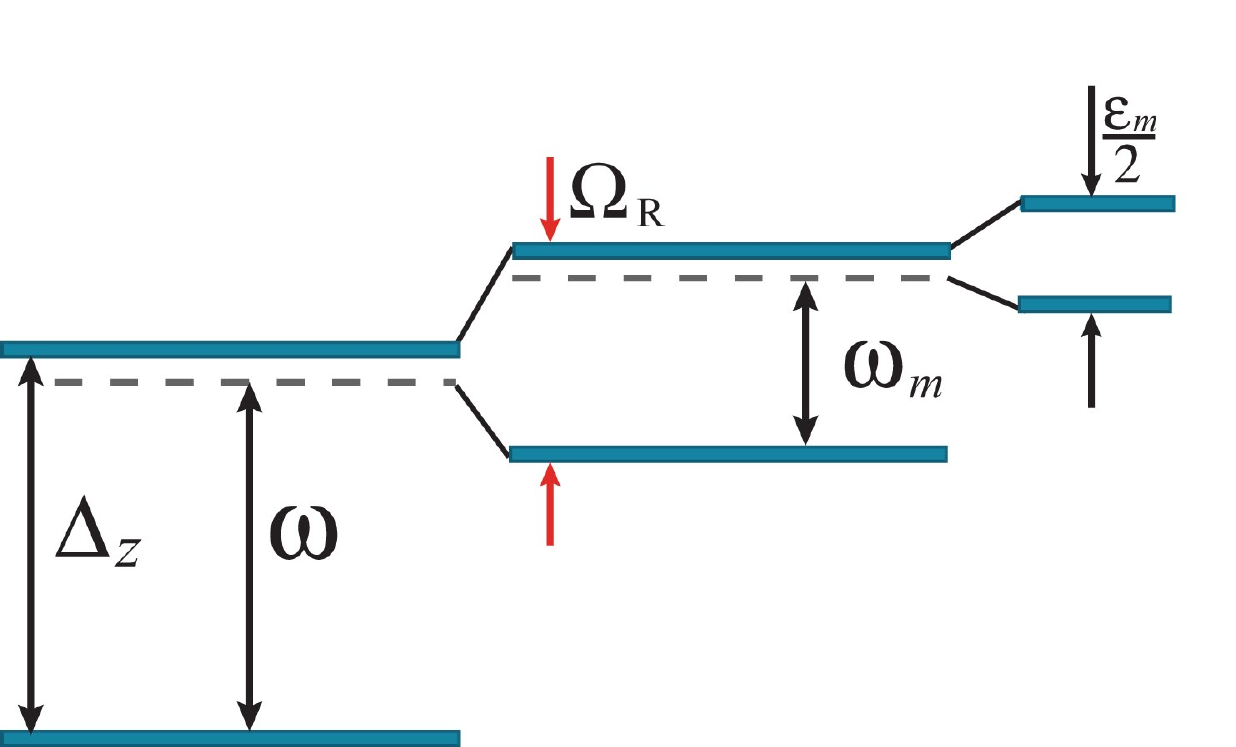}
\vspace{0.1cm}
\caption{Schematic illustration of  the second-order Rabi oscillations emerging
as a result of slow modulation of the static field, $B_{\s 0}$. Driving resonant ac field, splits the quasienergies by $\Omega_{\s R}$; modulation with frequency $\omega_{\s m}\approx \Omega_{\s R}$ leads to an additional splitting.
}
\label{fig:schematic}
\end{figure}

\subsection{Qualitative picture}
\label{Qualitative picture}
In Sect. II, based on the mapping Eq. (\ref{correspondence}), we concluded that Rabi oscillations are strongly sensitive to the modulation of
the longitudinal field when $\omega_{\s m}$ is close to $\Omega_{\s R}$. It is illustrative to discuss the effect of
modulation
near this condition using the language of
quasienergies. Within this language, without modulation,
the presence of a resonant field with frequency $\omega \approx \Delta_{{\scriptscriptstyle Z}}$
results in a degeneracy of quasinenergies
corresponding to Zeeman-split levels, see Fig. \ref{fig:schematic}.
Conventional Rabi oscillations emerge as a result of lifting this degeneracy (avoided crossing);
the magnitude of the splitting of quasienergies is $\Omega_{{\scriptscriptstyle R}}$.

With periodic modulation,
the Hamiltonian in the quasienergy representation remains time-dependent and the corresponding eigenstates are characterized
by {\em second-order}
quasienergies. When the frequency of modulation is close to
$\Omega_{{\scriptscriptstyle R}}$, the first-order Rabi-split
quasienergies become degenerate,  leading to their additional splitting, see Fig. \ref{fig:schematic}.
The magnitude of this additional splitting is controlled by the modulation amplitude. In the time domain, this additional splitting manifests itself via  {\em second-order} Rabi oscillations in the form of an envelope of the primary oscillations.

As we show in the next subsection, the shape of this envelope
is very sensitive to the detuning, $\omega_{\s m}-\Omega_{\s R}$,
of modulation frequency from the Rabi frequency, and to the detuning,
$\delta$, of the driving frequency from the Zeeman splitting, $\Delta_{\s Z}$.
\begin{figure}[h!]
\includegraphics[width=77mm]{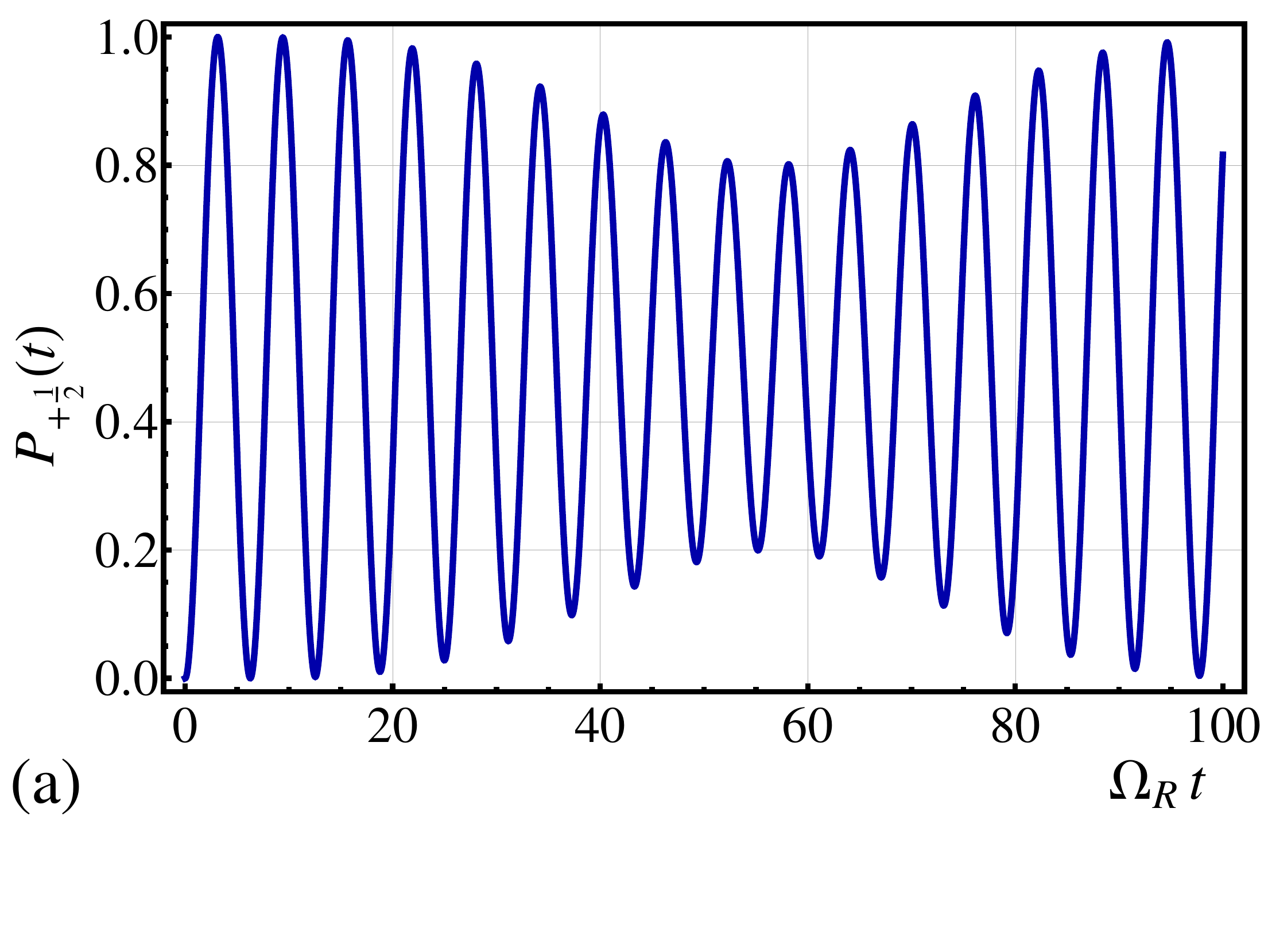}
\vspace{-0.6cm}\hspace{20mm}
\includegraphics[width=77mm]{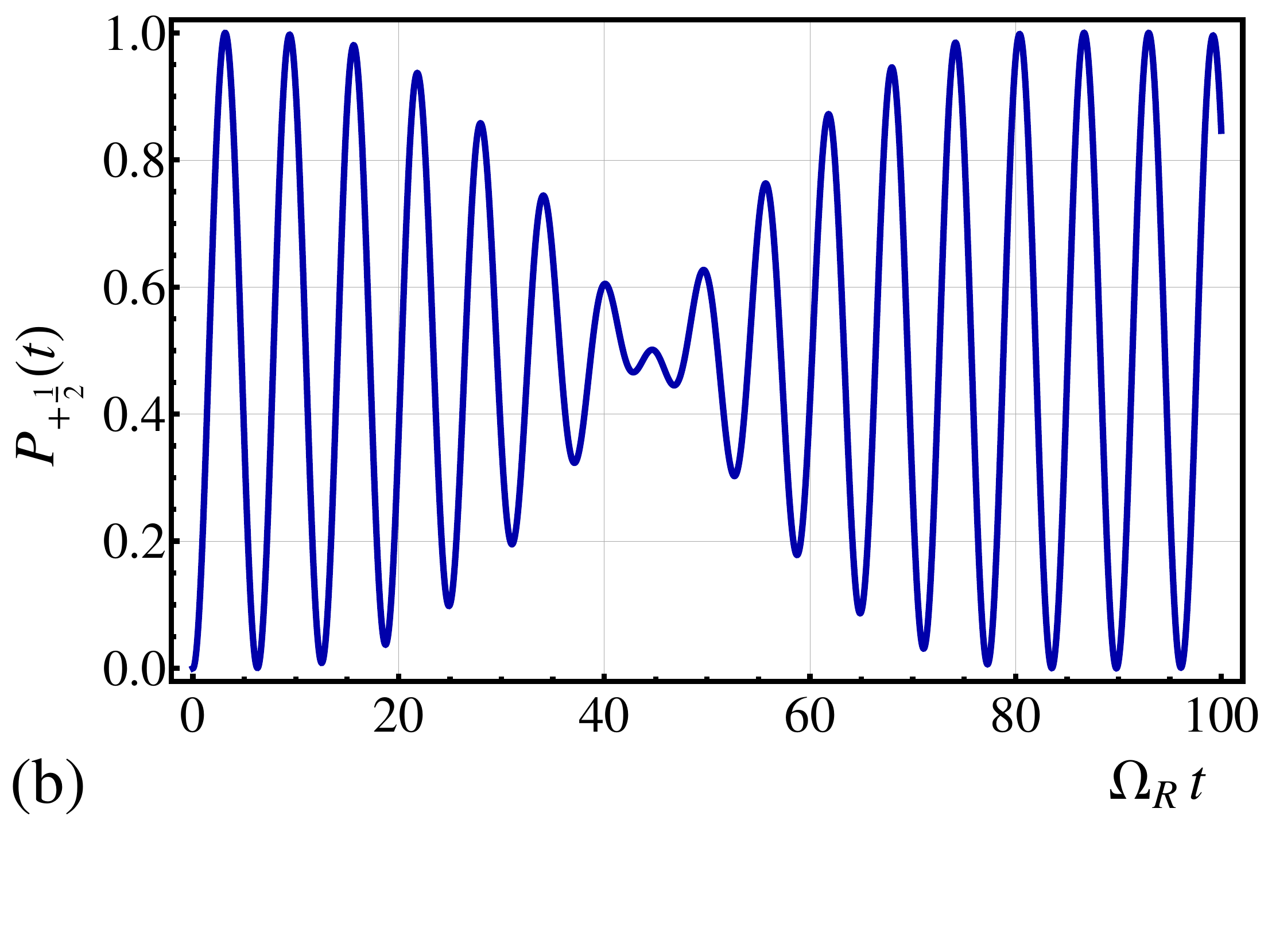}
\vspace{-0.6cm}\hspace{20mm}
\includegraphics[width=77mm]{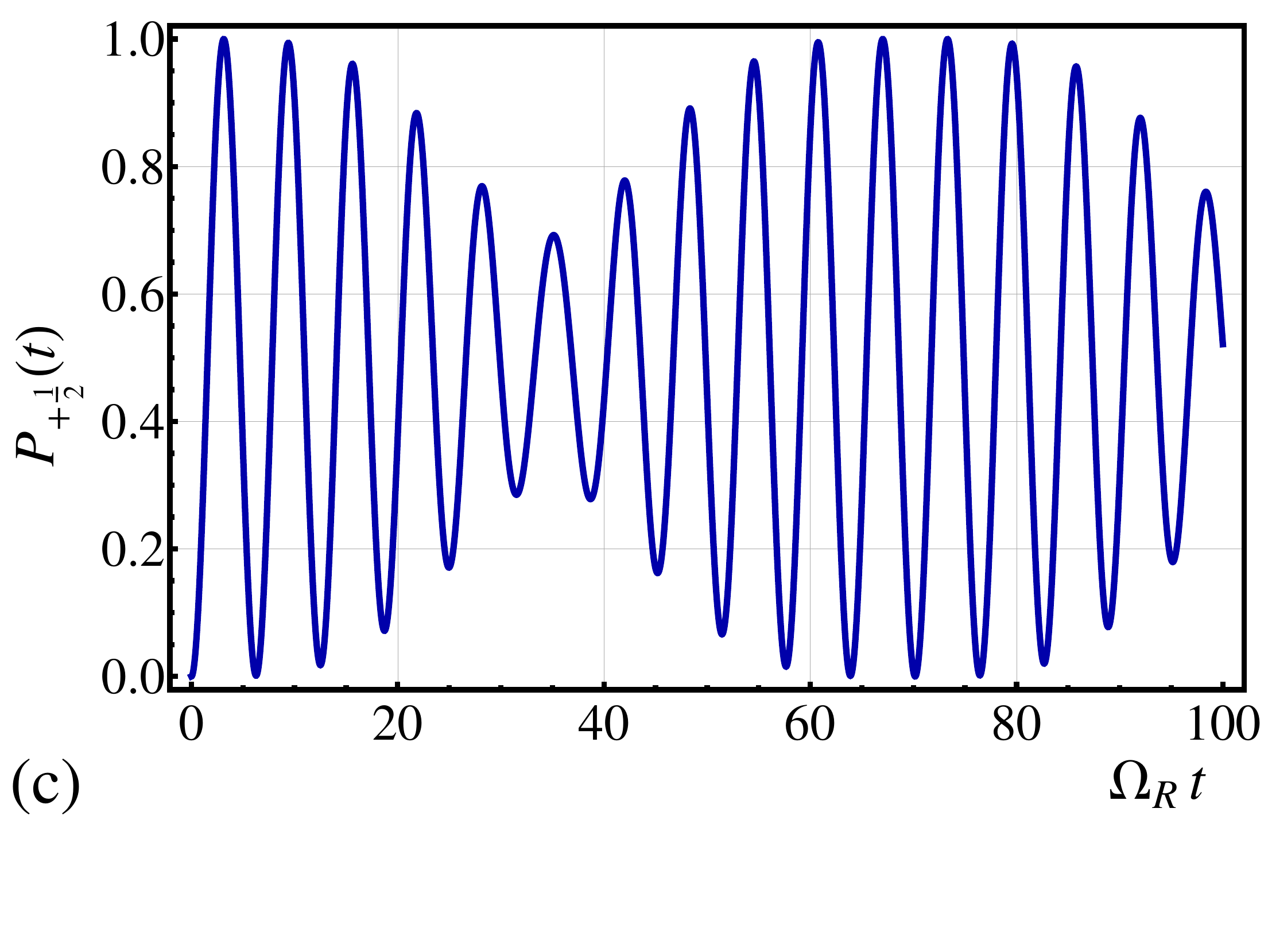}
\vspace{-0.5cm}
\caption{ Evolution of the Rabi oscillation pattern with
detuning of the modulation frequency from the Rabi frequency.
Spin dynamics, $P_{+\frac{1}{2}}(t)$, under the resonant drive, $\delta=0$, is  plotted from Eq. \eqref{zerodelta} for different
dimensionless detunings, $\kappa$:  $0.5$ (a),  $1.0$ (b), and $1.5$ (c).
The amplitude of modulation is $\varepsilon_{\s m}=0.1\,\Omega_{\s R}$.
Upon increasing $\kappa$ the envelope develops (a), reaches maximum (b), and gradually vanishes (c). }
\label{fig:nodetune}
\end{figure}


\subsection{The shape of the envelope}
\label{The shape of the envelope}
Near the condition $\omega_{\s m}=\Omega_{\s R}$    Eq. (\ref{secondorder}) can be solved in the resonant approximation, which amounts to keeping only two
terms in the Floquet expansion\cite{Shirley}:
\begin{equation}
\label{Floquet}
D_{+\frac{1}{2}}(t)=e^{i\lambda t}\Bigl(d_{{\scriptscriptstyle 0}}+d_{{\scriptscriptstyle -1}}
e^{-i\omega_{{\scriptscriptstyle m}} t}\Bigr),
\end{equation}
where $\lambda$ is the Floquet exponent. Substituting Eq. (\ref{Floquet}) into Eq. (\ref{secondorder})
and equating the resonant terms, we get the following
system of coupled equations for the
amplitudes $d_{{\scriptscriptstyle 0}}$, $d_{{\scriptscriptstyle -1}}$
\begin{eqnarray}
\label{system}
\left(\Omega_{{\s R}}^2+\delta^2+\frac{\varepsilon_{{\s m}}^2}{2}-4\lambda^2\right)
d_{{\s 0}}\!\!\!&=&\!\!\!-(\delta-\omega_{{\s m}})\varepsilon_{{\s m}}d_{{\s -1}},\hspace{1cm} \nonumber\\
\left(\Omega_{{\s R}}^2\!+\!\delta^2\!+\!\frac{\varepsilon_{{\s m}}^2}{2}-
4(\lambda-\omega_{{\s m}})^2\right)
d_{{\s -1}}\!\!\!&=&\!\!\!-(\delta+\omega_{{\s m}})\varepsilon_{{\s m}}d_{{\s 0}}.
\end{eqnarray}
For vanishing modulation, $\varepsilon_{{\s m}}\rightarrow 0$, both left-hand sides in the system
Eq. (\ref{system}) turn to zero at the degeneracy point $\lambda=\frac{1}{2}\omega_{{\s m}}$,
when the condition: $\omega_{{\s m}}=\omega_{{\s 0}}=\sqrt{\delta^2+\Omega_{{\s R}}^2}$ is met.
At finite $\varepsilon_{{\s m}}$ the system
yields the splitting of the Floquet exponents
\begin{equation}
\label{splitting}
\lambda_{{\s \pm}}=\frac{\omega_{{\s m}}}{2}
\pm
\frac{\varepsilon_{{\s m}}\,\Omega_{{\s R}}}{4\,\omega_{{\s 0}}}\sqrt{\kappa^2+1},
\end{equation}
where we introduced a dimensionless deviation
\begin{equation}
\label{kappa}
\kappa=\frac{2(\omega_{{\s m}}-\omega_{{\s 0}})\omega_{{\s 0}}}{\varepsilon_{{\s m}}\,\Omega_{{\s R}}}.
\end{equation}
of the modulation frequency from oscillation frequency, $\omega_{{\s 0}}$ and took
into account that $\vert \omega_{{\s m}}-\omega_{{\s 0}}\vert \ll \omega_{{\s 0}}$.
The steps leading from Eq. (\ref{Floquet}) to Eq. (\ref{splitting}) are the same as the steps
leading from Eq. (\ref{A+dotdot})   to the conventional Rabi oscillations Eq. (\ref{Rabi}).
The difference arises when one substitutes $\lambda_{{\s +}}$, $\lambda_{{\s -}}$ into
the system Eq. (\ref{system}), finds two linearly independent solutions, $D_{+\frac{1}{2}}^{{\s +}}(t)$
and $D_{+\frac{1}{2}}^{{\s -}}(t)$, and requires that their sum
satisfies the initial conditions Eq. (\ref{initialD}). After that, upon calculating the
occupation
 \begin{equation}
P_{+\frac{1}{2}}(t)=\left\vert D_{+\frac{1}{2}}^{{\s +}}(t)+D_{+\frac{1}{2}}^{{\s -}}(t)\right\vert^2,
\end{equation}
one arrives at the following generalization of Eq. (\ref{Rabi}) to the case of modulated longitudinal field

\begin{widetext}

\begin{align}
\label{finitedeltaHor}
&P_{+\frac{1}{2}}(t)=\frac{1}{\kappa^2+1}\Biggl[\left(1{\bf -}\frac{\delta(\kappa\,\Omega_{{\s R}}+\delta)}{\omega_{\s 0}^2}\right)\sin^2\!\left(\!\frac{\omega_{\s 0}}{2}+
\frac{\varepsilon_{\s m}\,\Omega_{\s R}}{4\,\omega_{\s 0}}\kappa\right)t
+\frac{\Omega_{\s R}(\kappa\,\Omega_{\s R}+\delta)}{2\,\omega_{\s 0}^2\big(\sqrt{\kappa^2+1}-\kappa\big)}
\sin^2\!\left(\!\frac{\omega_{\s 0}}{2}+\frac{\varepsilon{\s m}\,\Omega_{\s R}}{4\,\omega_{\s 0}}\left(\kappa-\sqrt{\kappa^2+1}\right)\right)t
\nonumber\\
&\hspace{1.5cm}+\frac{\delta(\kappa\,\Omega_{\s R}+\delta)}{\omega_{\s 0}^2}
\sin^2\!\left(\!\frac{\varepsilon_{\s m}\,\Omega_{\s R}}{4\,\omega_{\s 0}}\sqrt{\kappa^2+1}\!\right)t
-\frac{\Omega_{\s R}(\kappa\,\Omega_{\s R}+\delta)}{2\,\omega_{\s 0}^2\big(\sqrt{\kappa^2+1}+\kappa\big)}
\sin^2\!\left(\frac{\omega_{\s 0}}{2}+\frac{\varepsilon{\s m}\,\Omega_{\s R}}{4\,\omega_{\s 0}}\left(\kappa+\sqrt{\kappa^2+1}\right)\!\right)t
\Biggr].
\end{align}

\end{widetext}


We start the  analysis of  Eq. (\ref{finitedeltaHor}) from the case of zero detuning, $\delta=0$,
when it
simplifies considerably

\begin{eqnarray}
\label{zerodelta}
&&\hspace{-1cm}P_{+\frac{1}{2}}(t)=
\frac{1}{\kappa^2+1}\Biggl[\sin^2\!\left(\frac{\Omega_{\s R}}{2}+
\frac{\varepsilon_{\s m}}{4}\kappa\!\right)\!t
\nonumber\\
&&\hspace{-1cm}+\frac{\kappa}{2\big(\!\sqrt{\kappa^2+1}-\kappa\big)}\sin^2\!\left(\!\frac{\Omega_{{\s R}}}{2}\!+\!\frac{\varepsilon_{\s m}}{4}\!\left(\!\kappa-\sqrt{\kappa^2+1}\!\right)\!\right)\!t
\nonumber\\
&&\hspace{-1cm}-\frac{\kappa}{2\big(\!\sqrt{\kappa^2+1}+\kappa\big)}\sin^2\!\left(\frac{\Omega_{{\s R}}}{2}\!+\!\frac{\varepsilon_{\s m}}{4}\!\left(\!\kappa+\!\sqrt{\kappa^2+1}\right)\!\right)\!t
\Biggr]\!.\vspace{-1cm}
\end{eqnarray}
This expression is consistent with Ref. \onlinecite{Saiko1} where a related quantity, namely the absorption of an external field by a resonantly driven spin-$\frac{1}{2}$ system in a modulated longitudinal field, has been studied.

The last two terms in Eq. (\ref{zerodelta}) describe the second-order Rabi oscillations since their frequencies
are split in agreement with qualitative picture Fig. \ref{fig:schematic}.
Naturally, taking the limit $\varepsilon_{\s m}\rightarrow 0$,
we recover the conventional Rabi oscillations,  $P_{+\frac{1}{2}}(t)=\sin^2\Bigl(\frac{\Omega_{\s R}t}{2}\Bigr)$, for any $\kappa$. A nontrivial feature of Eq. (\ref{zerodelta}) is that exactly at $\omega_{\s m} =\Omega_{\s R}$, when $\kappa=0$, the last two terms cancel each other, i.e., the Rabi oscillations are {\em unaffected} by the modulation.
Gradual development of the second-order oscillations upon increasing $\kappa$ is illustrated in Fig.~\ref{fig:nodetune}. We see that the effect of
modulation of longitudinal field is most pronounced
at dimensionless detuning $\kappa \approx 1$, where the modulation of the Rabi oscillations is complete, see Fig.~\ref{fig:nodetune}b.
Upon further increasing $\kappa$, the effect of modulation vanishes above $\kappa =4$.
\begin{figure}
\includegraphics[width=77mm]{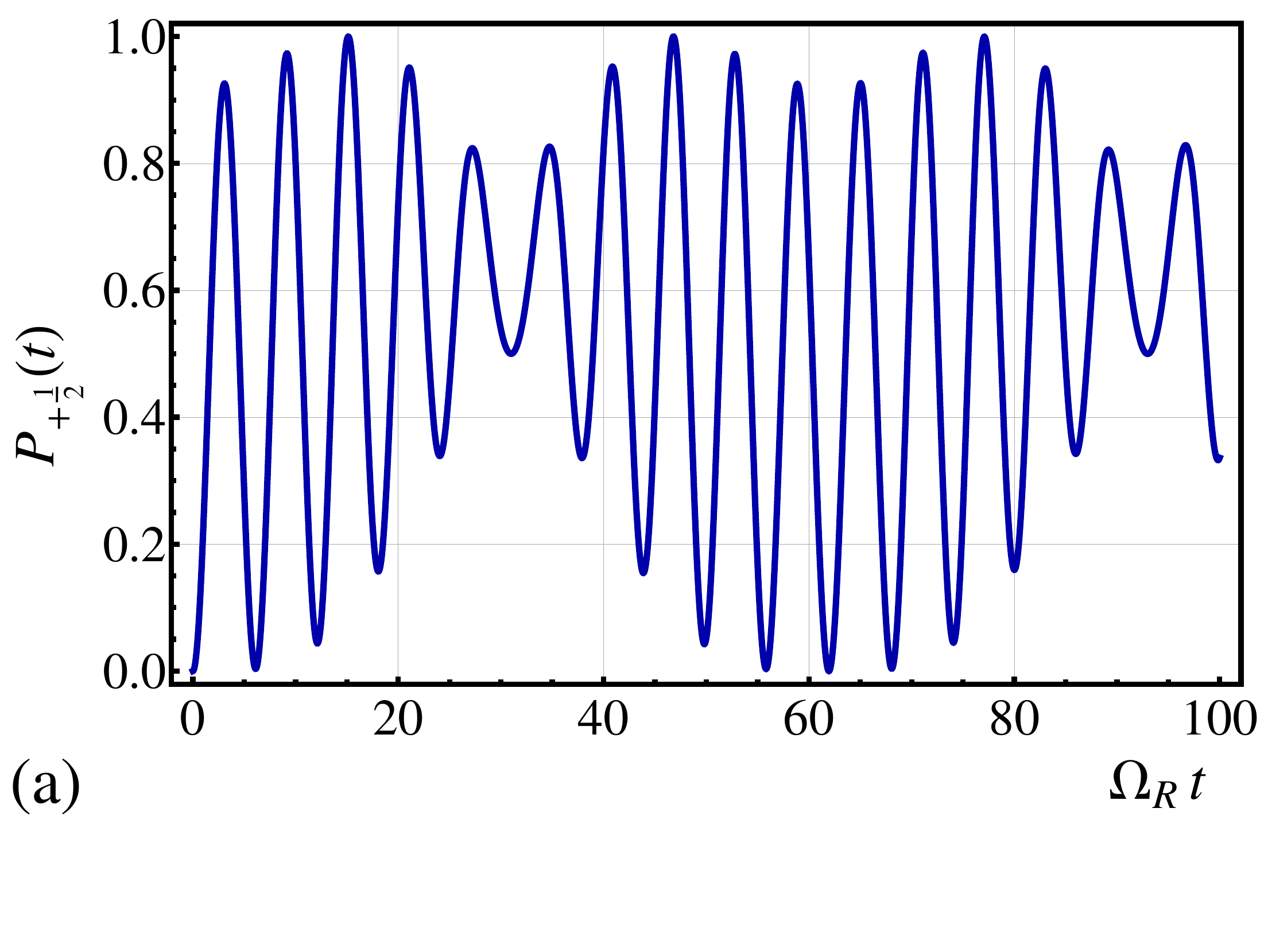}
\includegraphics[width=77mm]{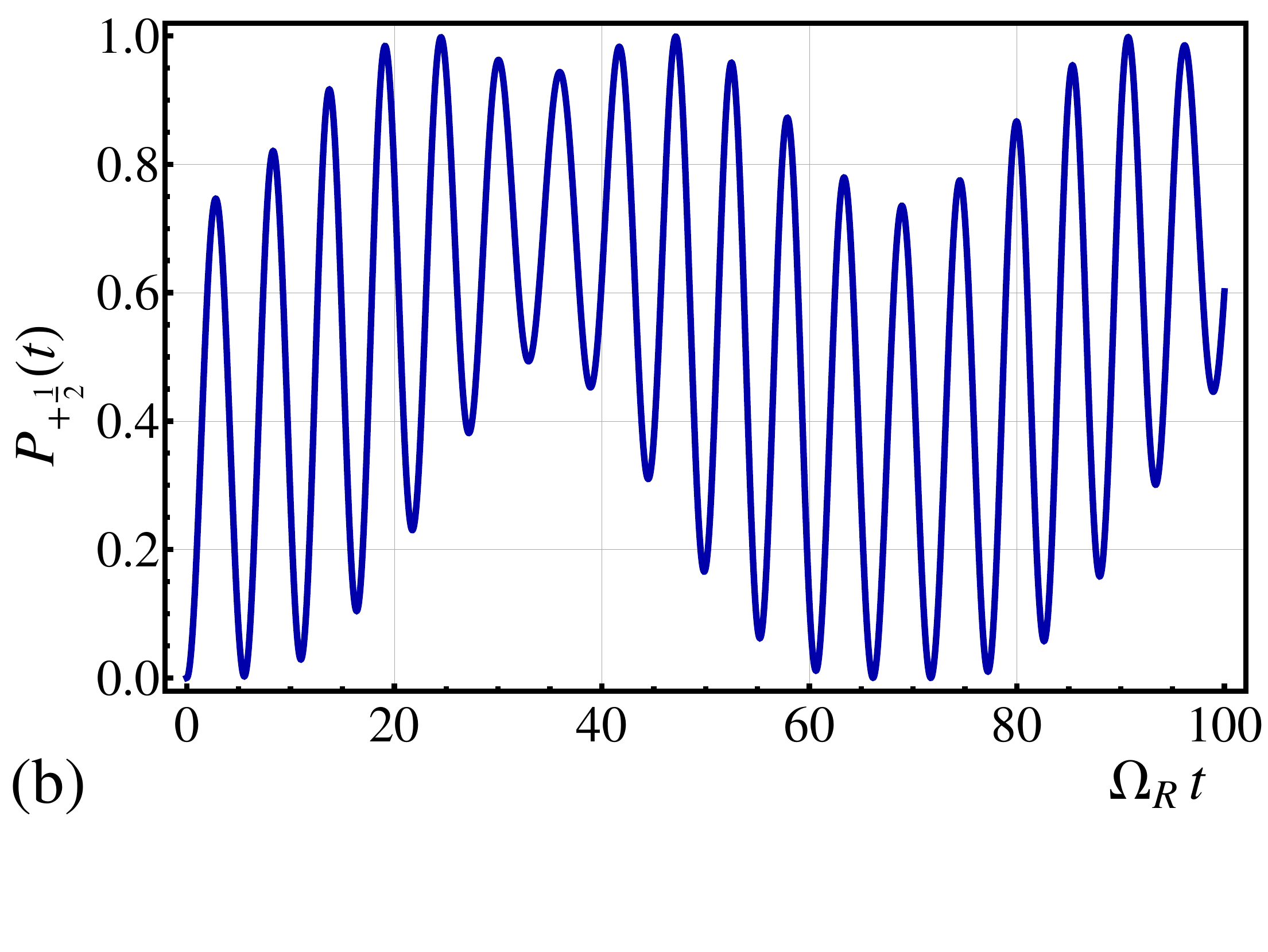}
\includegraphics[width=77mm]{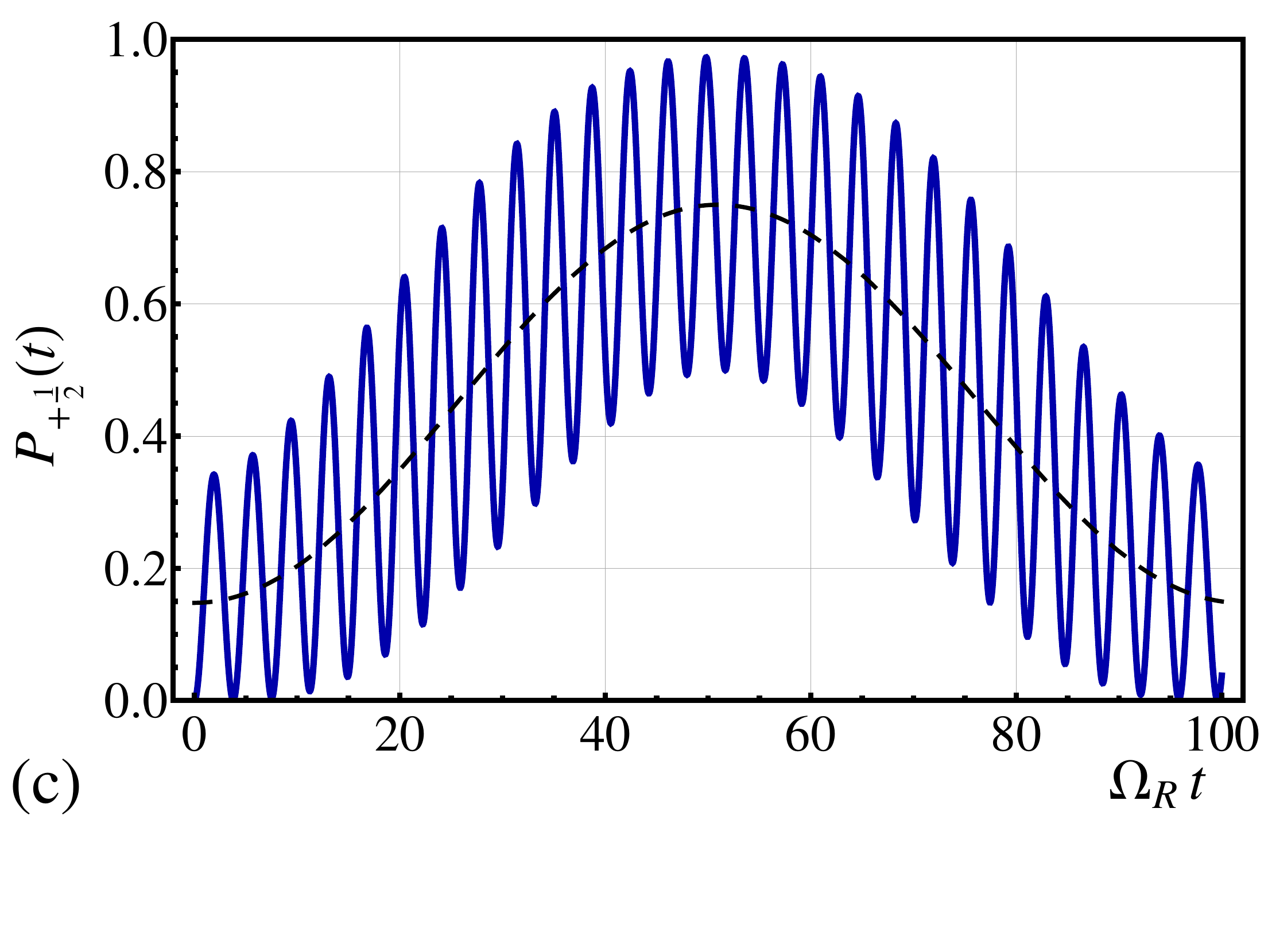}
\vspace{-0.3cm}
\caption{Regime of weak resonant modulation. Evolution of the Rabi oscillation pattern with detuning, $\delta$, of the driving frequency, $\omega$, from the level splitting, $\Delta_{\s Z}$.
Spin dynamics, $P_{+\frac{1}{2}}(t)$, is plotted from Eq. \eqref{finitedeltaHor} for three positive values of
$\frac{\delta}{\Omega_{\s R}}$: $0.3$ (a), $0.6$ (b), and $1.4$ (c).
The amplitude of modulation is $\varepsilon_{\s m}=0.15\,\Omega_{\s R}$;
dimensionless detuning of the modulation frequency from $\Omega_{\s R}$
is $\kappa=1.0$. The pattern gradually evolves from envelope of the Rabi oscillations (a) to weak oscillations around the modulated baseline (c).}
\label{fig:detune_pos}
\end{figure}

Beyond the resonant approximation\cite{Saiko1},
the difference, $\omega_{\s m}-\omega_{\s 0}$,
in Eq. (\ref{kappa}) acquires a correction
$\sim \varepsilon_{\s m}^2/\Omega_{\s R}$,
which is an analog of the Bloch-Siegert shift\cite{BlochSiegert}.

Solving Eq. (\ref{secondorder}) in the resonant approximation is also
permitted when $\omega_{\s m}$ is close to $\omega^{(p)}_{\s m}= \frac{\Omega_{\s R}}{2p+1}$. Then the two resonating terms in the Floquet expansion are
$d_{{\scriptscriptstyle 0}}$ and $d_{\s -(2p+1)}$. The
splitting of the Floquet exponents, at the degeneracy point, $\lambda=\frac{\Omega_{{\scriptscriptstyle R}}}{2(2p+1)}$,
takes place in the ($2p+1$)-th order in parameter $\varepsilon_{{\scriptscriptstyle m}}/\Omega_{{\scriptscriptstyle R}}$. The magnitude of the splitting can
be obtained by readjusting notations in the expression for the width of multiphoton resonance obtained in Ref. \onlinecite{Shirley}
\begin{eqnarray}
\label{fromshirley}
\frac{\varepsilon_{\s m}^{(p)}}{2}=\Bigl(\frac {\varepsilon_{\s m}}{4}\Bigr)\Biggl(\frac
{\varepsilon_{\s m}}{\omega^{(p)}_{\s m}}\Biggr)^{2p}\frac{1}{2^{2p-1}(p!)^2}.
\end{eqnarray}
In the domain $(\omega_{\s m}-\omega^{(p)})\sim \varepsilon_{\s m}^{(p)}$ Eq. (\ref{zerodelta}) remains
valid
 upon the replacement
$\varepsilon_{\s m} \rightarrow \varepsilon_{\s m}^{(p)}$ and describes second-order multiphoton Rabi oscillations. Naturally, if the modulation
is not purely sinusoidal but contains a harmonics,
\begin{figure}[h!]
\includegraphics[width=78mm]{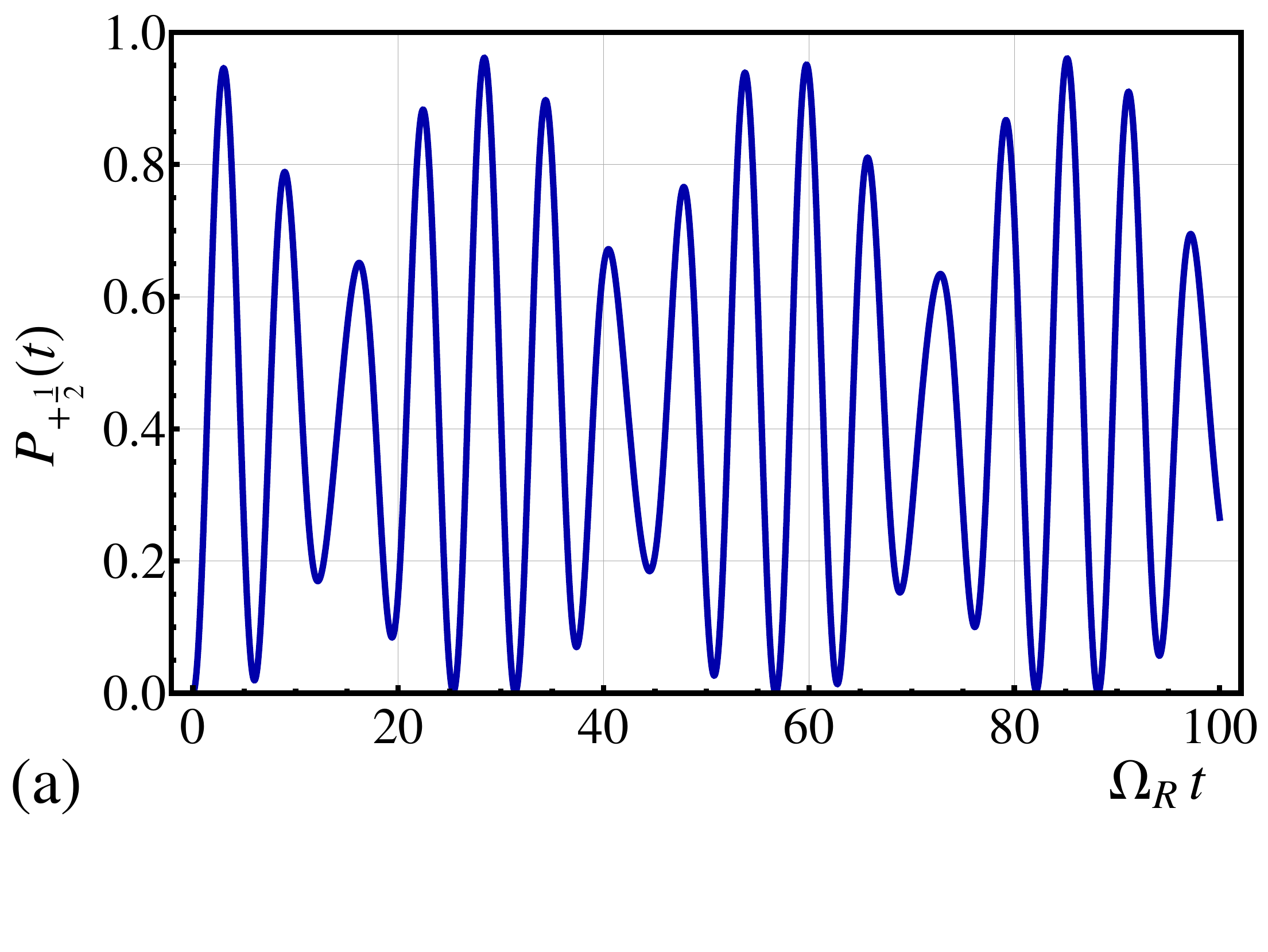}
\vspace{-0.6cm}
\includegraphics[width=78mm]{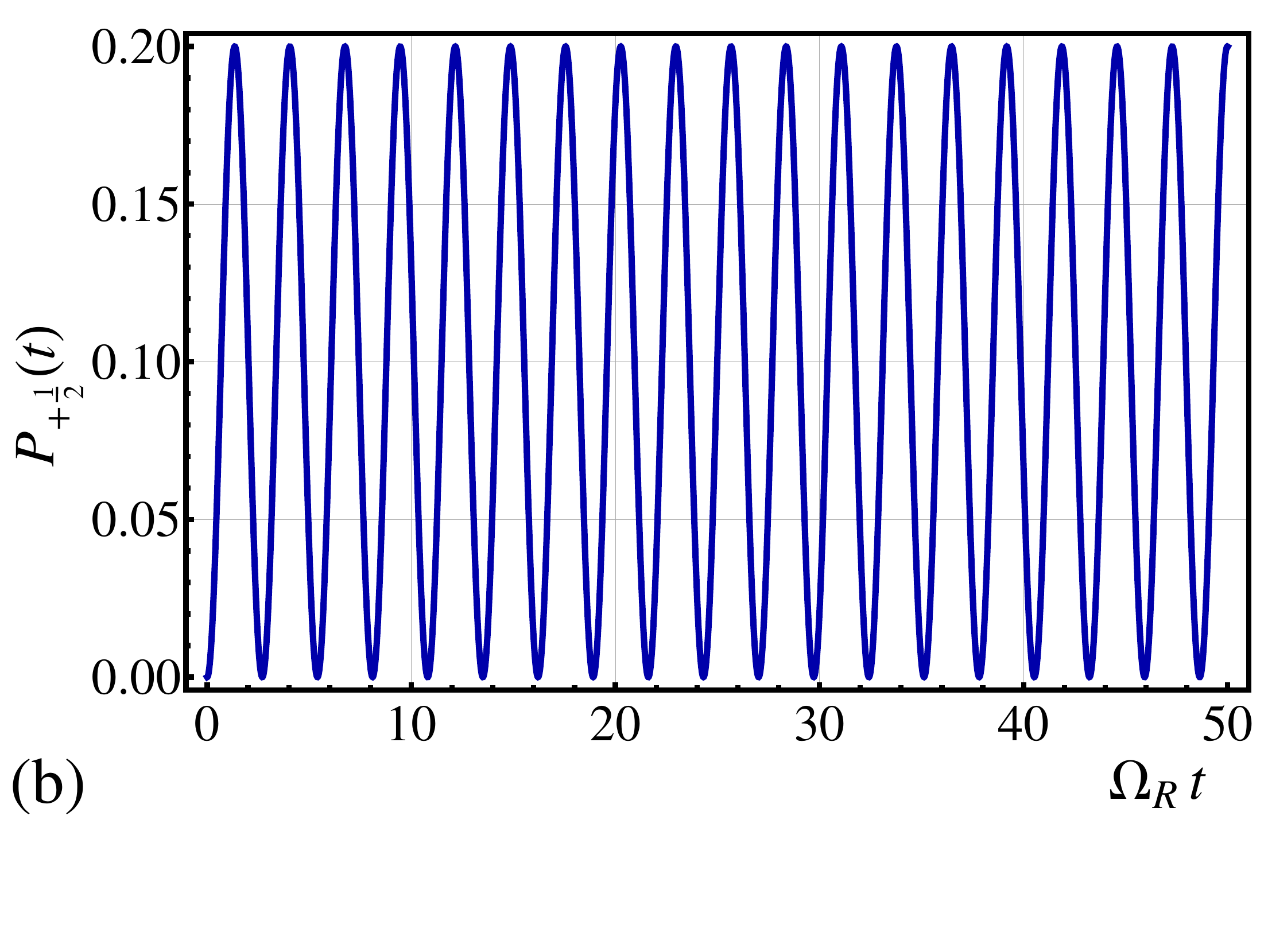}
\includegraphics[width=78mm]{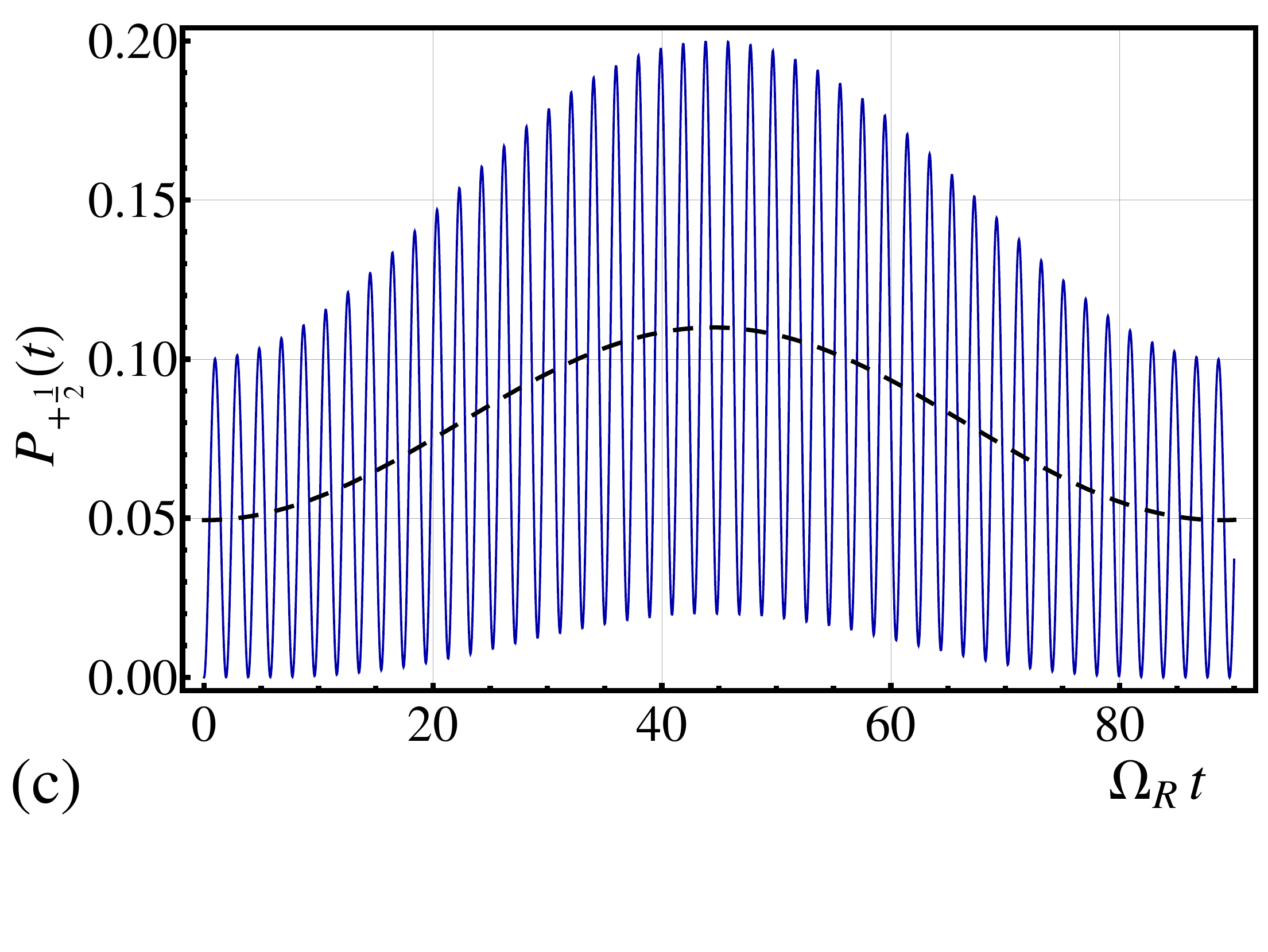}
\vspace{-0.3cm}
\caption{Same as Fig.~\ref{fig:detune_pos}
for three negative values of
$\frac{\delta}{\Omega_{\s R}}$: $-0.2$ (a), $-2$ (b), and $-3$ (c).
The pattern evolves from envelope (a) to homogeneous oscillations (b).
Weak modulation of the baseline emerges only at large detuning (c). }
\label{fig:detune_neg}
\end{figure}
say $l$,  the envelope of the Rabi oscillations near $\omega_{\s m}=\Omega_{\s R}/l$ will already develop  in the lowest order.




In the absence of modulation, finite  detuning, $\delta$, of the driving frequency from the resonant
frequency, $\Delta_{\s Z}$, leads to the reduction of the
period of the Rabi oscillations and to the reduction
of their amplitude, see Eqs. (\ref{Rabi}), (\ref{frequency}). Remarkably, in the presence of
modulation, near the condition $\omega_{\s m}=\Omega_{\s R}$, finite $\delta$ causes a new {\em qualitative}
feature in the shape of the Rabi oscillations. Moreover, the
effect of finite $\delta$ depends strongly on its {\em sign}.

In Fig.~\ref{fig:detune_pos} we plot from
Eq. (\ref{finitedeltaHor}) the evolution
of the Rabi-oscillation pattern as $\delta$
increases in the positive direction, while parameter
$\kappa$ is kept constant. We see that, while for small enough $\delta$, the prime effect of modulation is still a slow envelope of the oscillations, upon further increasing of $\delta$, the envelope gradually transforms into the oscillations of the {\em baseline}. At large enough $\delta$, Fig. ~\ref{fig:detune_pos}c, these
oscillations of the baseline
dominate the Rabi-oscillation pattern.

As $\delta$ increases in the negative direction,
see Fig.~\ref{fig:detune_neg}, the pattern of the Rabi oscillations
evolves differently. At small $|\delta|$ modulation
gives rise to the envelope, similar to the case of small positive
$\delta$. However, at a certain $|\delta|$, see Fig ~\ref{fig:detune_pos}b, the magnitude of the oscillations
does not change with time at all,
suggesting that the effects of two detunings, $\omega_{\s m}-\omega_{\s 0}$, and, $\omega-\Delta_{\s Z}$, {\em cancel each other}. Only upon further increase of $|\delta|$ the magnitude of oscillations
decreases significantly, and they proceed around the
slowly oscillating baseline, as shown in Fig. ~\ref{fig:detune_pos}c. The plots in Figs. \ref{fig:detune_pos}, \ref{fig:detune_neg} are drawn for
the ratio $\varepsilon_{\s m}/\Omega_{\s R} = 0.15$ and $\varepsilon_{\s m}/\Omega_{\s R} = 0.2$, respectively, however  general features of the modulation pattern depend weakly on this ratio.

\section{Testing theory with NMR Experiment}
\begin{figure}[b]
\includegraphics{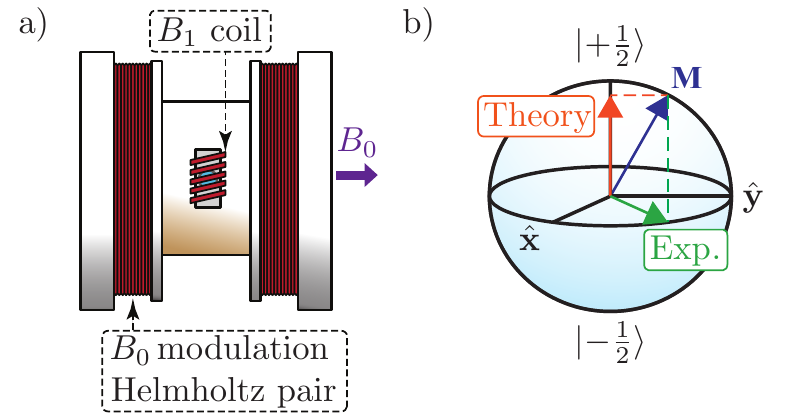}
\caption{ \label{fig:coil_scheme}(Color online) a) A schematic of the NMR probe used in the experiments. A traditional NMR coil ($B_{\s 1}$) is accompanied by a $ B_{\s 0}$ modulation Helmholtz pair that is coaxial with the $ B_{\s 0}$ field. b) Graphical description of the relation between theoretical predictions and experimental NMR data. The theoretical predictions have the magnetization $\mathbf{M}$ projected onto the $z$-axis of the Bloch sphere. The experimental data is the projection of the magnetization onto the $xy$-plane of the Bloch sphere. 
Eq. \eqref{experiment} relates the theory to experiment. }
\end{figure}

The introduction of the longitudinal modulation field
$\varepsilon(t)$ to investigate the general behavior of Rabi
oscillations is highly compatible with conventional experimental NMR
methods of inductive detection in thermally generated spin
ensembles\cite{Book2}. For the large $ B_{\s 0}$ field generally required to achieve
sufficient signal-to-noise ratio (SNR), the condition $B_{\s 1}\ll  B_{\s 0}$
cannot be avoided, but the additional small-amplitude modulation
field is easily realized (Fig.~\ref{fig:coil_scheme}a). The three regimes elucidated in
Sect.~\ref{Fast Modulation}, \ref{Strong Modulation}, and
\ref{Weak-resonant Modulation} were thus explored using
straightforward NMR of protons in water (gyromagnetic ratio $\gamma
= g\mu_N = 4.25775$~kHz/G, where $\mu_N$ is the nuclear magneton).
We needed only to take some additional care to ensure a stable and
highly homogeneous rf driving field $B_{\s 1}$ across the sample, so that
the pattern of Rabi oscillations could be observed over many
periods.

All experiments were performed in a horizontal-bore 2-Tesla
superconducting magnet (Oxford Instruments). A conventional solenoidal
single-coil transmit/recieve probe (5 turns, 1 cm diam and 2.5 cm
long) was series-tuned with a capacitor to the proton Zeeman
resonance at 88.8~MHz. A 50-$\Omega$ resistor in series with these
elements provided a matching impedance to the transmit and receive
amplifiers. This ``low-Q" probe sacrifices some SNR for a robust and
flat frequency response that is negligibly affected by the
accompanying modulation field\cite{Book2,Book3}. The water sample is centered in and
occupies about 25\% of the coil volume; it is contained in small
tube made of PTFE (Teflon$^{\rm TM}$), which minimizes the
magnetic-susceptibility difference between the sample and its
immediate surroundings. Both the low filling factor and PTFE tube
serve to enhance the $B_{\s 1}$-field homogeneity across the sample. The
modulation field was provided by a 5-cm-radius Helmholtz pair
(coaxial with the main $ B_{\s 0}$ field) wound on a form that has the
probe coil at its center (see Fig.~\ref{fig:coil_scheme}a). A NMR spectrometer (Tecmag Redstone, model HF2-1RX)  with two independent transmission channels was used to
transmit pulses to both the 88.8~MHz proton probe and the 0-100~kHz
modulation coils, and subsequently to acquire and digitize the
free-induction-decay (FID) signal generated in the probe coil. The
$B_{\s 1}$ rf pulse was amplified with a 2~kW amplifier (Tomco model BTO2000-AlphaSA)
conventionally designed for solid-state NMR, but whose high output
power allowed the coherent nutation of the proton spins through many
Rabi-oscillation periods in a relatively short pulse time. The
$ B_{\s 0}$-modulation pulse was provided by a DC-50~kHz amplifier
normally used for gradient coils in imaging applications (Techron
AN7780); the inductance of the Helmholtz pair was matched to the
gradient amplifier's output specifications in order to minimize
ring-down and associated cross-talk to the NMR coil.

\begin{figure}
\includegraphics{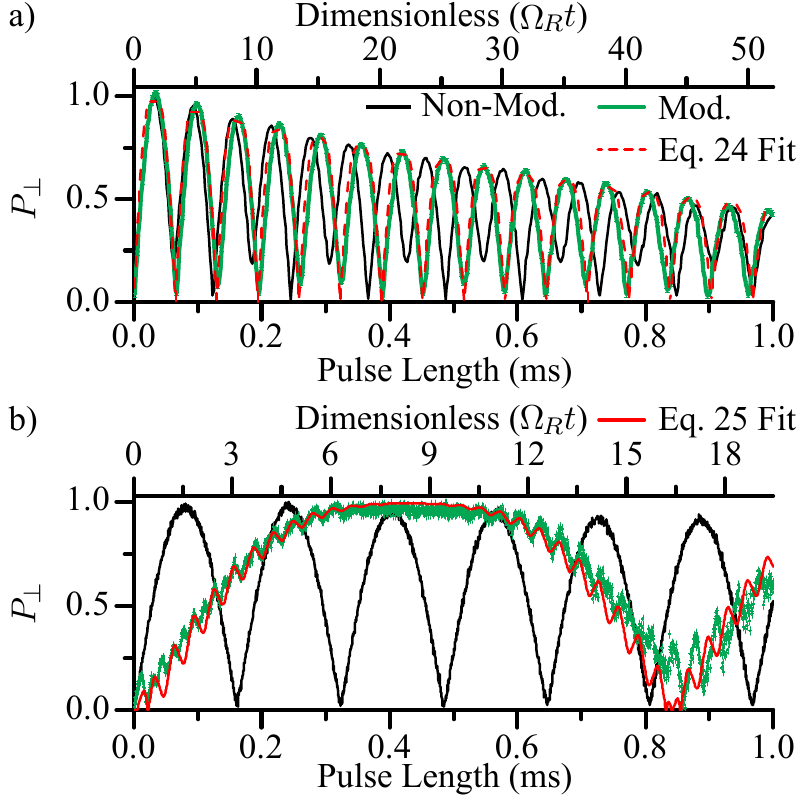}
\caption{\label{fig:exp_fastmod} (Color online) Rabi oscillations in the fast-modulation regime (green) compared to non-modulated data (black).  Transverse spin projection $P_{\perp}$ is plotted vs. pulse length; the latter is linearly related to the nutation angle. a) The modulation amplitude $\varepsilon_{\s m}$ is small compared with $\omega_{\s m}$, and the modulated data show the expected small upward shift in the oscillation frequency. A fit of the modulated data to Eq.~\eqref{Modiefied} is shown by the dashed red line. The experimentally determined values of the modulation frequency $\omega_{\s m}/2\pi\! =\! 41.30  \pm 0.01$~kHz, modulation amplitude $\varepsilon_{\s m}/ 2\pi\gamma \! =\!4.96 \pm 0.06$~G, and the Rabi frequency $\Omega_{\s R}/2\pi\! = \!8.26 \pm 0.05$ kHz (determined from the non-modulated data) are input as fixed parameters. The fit includes multiplication of Eq.~\eqref{Modiefied} by $e^{-t/{T_{\s 2}^*}}$ to account for the overall decay; the fit then yields $T_{\s 2}^* = 1.28$ $\pm 0.02$ ms.  b) The modulation amplitude is large compared with $\omega_{\s m}$, and the oscillation frequency decreases significantly, in accordance with Eq.~\eqref{Modiefied}.  Here, $\omega_{\s m}/2\pi\! =\! 31.5  \pm 0.01$~kHz and $\Omega_{\s R}/2\pi\! = \!3.10 \pm 0.02$ kHz are experimentally determined. A fit that incorporates the correction of Eq.~\eqref{correction} is shown in red, from which $\varepsilon_{\s m}/ 2\pi\gamma \! =\!13.0 \pm 0.1$~G is extracted; larger values of $\varepsilon_{\s m}$ were difficult to measure experimentally.
The correction gives rise to the oscillations at $\omega_m$ that ride on top of the much slower Rabi oscillations.
The weaker $B_{\s 1}$ produces negligible $T_{\s 2}^*$ decay.}
\end{figure}

\begin{figure}[t]
\includegraphics{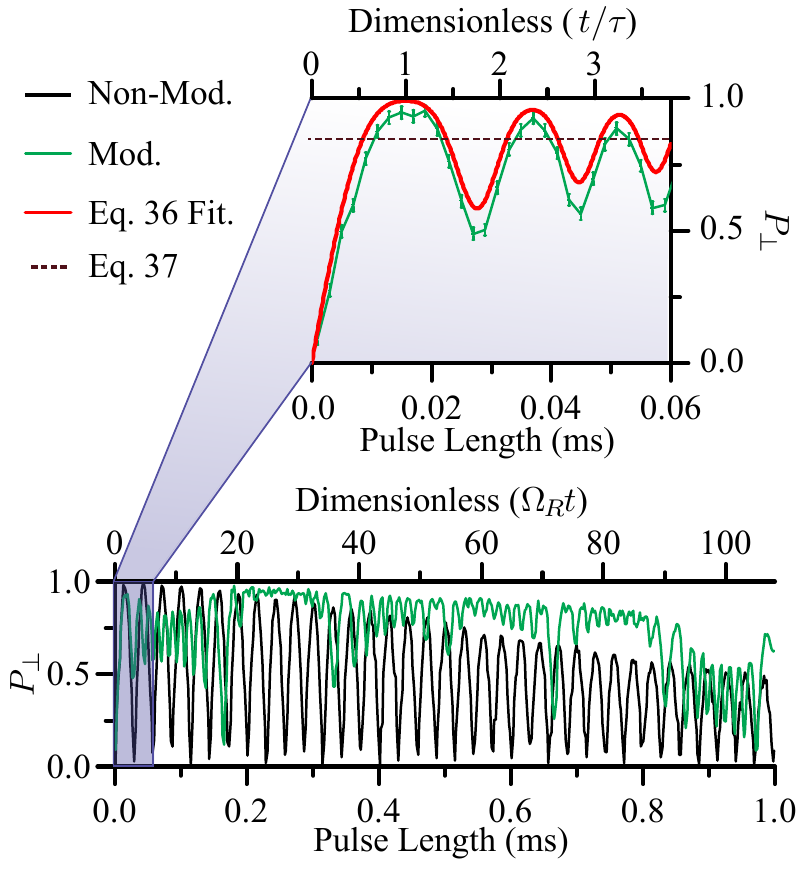}
\caption{\label{fig:exp_slowstrong}(Color online) Rabi oscillations in the strong-modulation regime (green) compared to non-modulated data (black).  Transverse spin projection $P_{\perp}$ is plotted vs. pulse length; the latter is linearly related to the nutation angle. The lower part of the figure shows the complete data set. Data without modulation (black) yields a Rabi frequency $\Omega_{\s R}/2\pi \!=\! 17.7 \pm 0.1$ kHz. Data with modulation are shown in
green; the applied modulation frequency was $\omega_{\s m}/2\pi \!=\! 3.0 \pm 0.1$ kHz. The critical regions near $\varepsilon(t) = 0$ occur every $0.166$ ms and are characterized by non-trivial oscillation behavior; the first such region is shown expanded in the upper part of the figure. These early-time data show the decreasing oscillation period characteristic of the parabolic cylinder functions. The results of a fit (red) of the early-time data to Eq.~\eqref{probability} yields $\varepsilon_{\s m}/2\pi\gamma\ \!=\! 7.0 \pm 0.05$ G, which then leads to $|\nu| \!=\! 0.9 \pm 0.1$. An
asymptote (dashed maroon) is calculated from Eq.~\eqref{probability} and is also shown.  Well away from the zero-crossings of $\varepsilon(t)$, the magnetization continues to nutate in a more regular fashion; that it nutates at all is the result of not sufficiently satisfying the limit $\varepsilon_{\s m} \gg \Omega_{\s R}$.}
\end{figure}

\begin{figure*}
\includegraphics{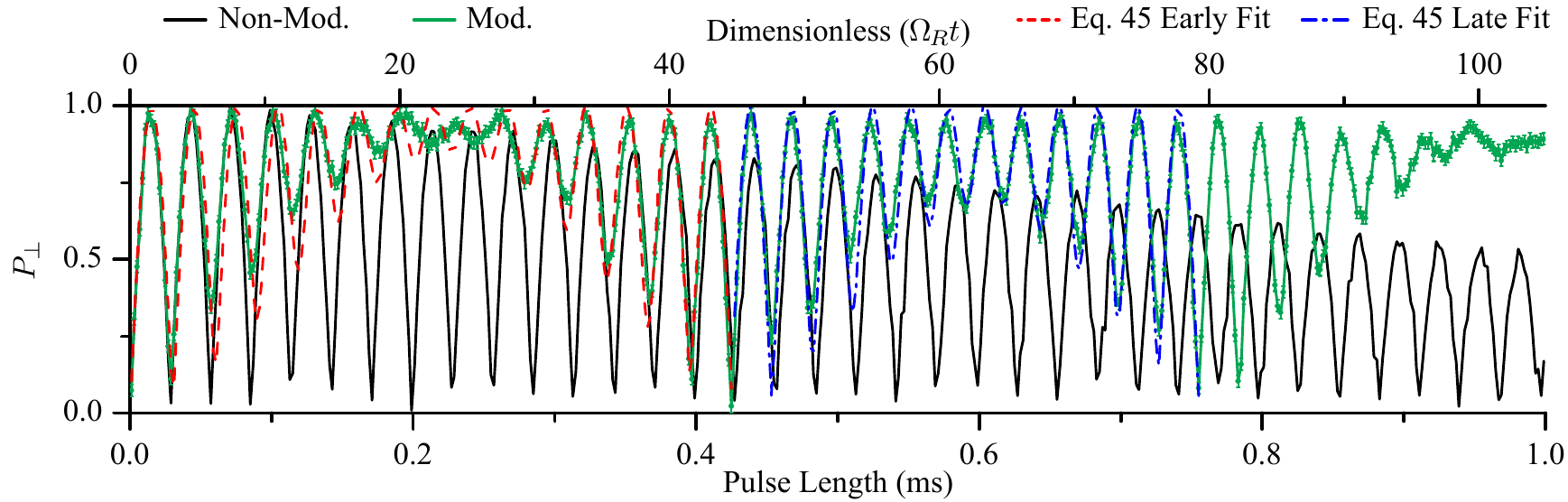}
\caption{\label{fig:exp_slow_resonant} (Color online)
Rabi oscillations in the regime of weak resonant modulation, showing definite beats that depend on the parameter $\kappa$. The Rabi frequency is found from the non-modulated data to be $\Omega_{\s R} /2\pi \! = \!16.7 \!\pm \! 0.7$ kHz. The sensitivity of the fit to small changes in the modulation frequency $\omega_m$ and experimental limitations led us to float most of the experimental parameters and to perform two separate fits of these data to Eq.~\eqref{zerodelta} for the early-time (red dashed) and late-time (blue-dashed) portions of the overall acquisition. For the early-time fit, the extracted values are: $\Omega_{\s R}/2\pi \! =\! 16.5 \!\pm \!0.2$ kHz, $\varepsilon_{\s m}/2\pi\gamma\! =\! 0.91 \pm 0.04$ G, and $\omega_{\s m}/2\pi \!= \!17.56 \! \pm \! 0.05$, yielding $\kappa = 0.54 \pm 0.14$. for the late-time fit, the extracted values are: $\Omega_{\s R}/2\pi \! =\! 17.0 \!\pm \!0.2$ kHz, $\varepsilon_{\s m}/2\pi\gamma \! =\! 1.2 \pm 0.1$ G, and $\omega_{\s m}/2\pi \!= \!18.03 \! \pm \! 0.09$, yielding $\kappa = 0.40 \pm 0.12$. The data exhibit the beating envelope characteristic of this modulation regime.}
\end{figure*}

The FID signals were acquired on resonance ($\delta = 0$) in
single-shot fashion (no signal averaging), and the FID signal
strength was plotted vs. $B_{\s 1}$ pulse length. Each data point
displayed is the result of examining the corresponding FID to
determine the magnitude of the transverse magnetization at a fixed
time delay
$\approx 0.2$~ms
from the end of the $B_{\s 1}$ rf pulse.
The $B_{\s 1}$ and modulation pulses are essentially applied
simultaneously, with the modulation pulse nominally starting at
$\varepsilon(t)= 0$ just as the $B_{\s 1}$ pulse starts. We assume that
the $B_{\s 1}$-pulse amplitude is fixed and the pulse length is linearly
related to the nutation angle of the proton spins. Because of slight
transient changes in power delivered by the rf amplifier at the very
beginning of the $B_{\s 1}$ pulse (minimized by using a low-Q probe),
this linear relationship is most accurately
observed for longer pulse times. Indeed, when we determine
$\Omega_{\s R}$ experimentally using non-modulated data, we rely on
measuring the average period of the later-time oscillations. With
this caveat in mind, the plots we show below represent Rabi
oscillations in the three modulation regimes of interest. To decrease
the overall data acquisition time, the water sample contained
dissolved copper sulfate (CuSO$_4$) to reduce the longitudinal
relaxation time $T_{\s 1}$ to $\approx 100$~ms. One must
generally wait at least several times $T_{\s 1}$ between FID acquisitions
to allow the magnetization to recover to its characteristic
thermal-equilibrium value. Intrinsic transverse relaxation
(characterized by $T_{\s 2}$) is typically on the order of $T_{\s 1}$ in
weakly interacting liquids. We observed transverse decoherence
caused by residual inhomogeneity (characterized by $T_{\s 2}^*<T_{\s 2}$) in
the $B_{\s 1}$ field, which manifests most clearly in our data as the
overall decay of the non-modulated Rabi oscillations after many
characteristic periods (see Figs.~\ref{fig:exp_fastmod}-\ref{fig:exp_slow_resonant}).

We note that the predictions made in Sections~\ref{Fast
Modulation}-\ref{Weak-resonant Modulation} are formulated in terms
of the projection of the magnetization onto the $z$-axis of the Bloch
sphere (see Fig.~\ref{fig:coil_scheme}b). Rabi oscillations occur
between the high- and low-energy states defined by $ B_{\s 0}$, where the
low-energy state has magnetization parallel to $ B_{\s 0}$. However, the
observable in a conventional NMR experiment is the projection of the
magnetization onto the
$xy$-plane
of the Bloch sphere. Noting also that the initial conditions at time
$t=0$ for our predictions have $P_{+\frac{1}{2}}(0) = 0$, whereas
the experiments have $P_{+\frac{1}{2}}(0) = 1$, a simple
transformation of the
$z$-axis prediction to the
$xy$-plane can be accomplished by
\begin{equation}
\label{experiment}
P_{\perp} = 
 \sin\big[\arccos (2P_{+\frac{1}{2}} -1)\big] = 2\sqrt{P_{+\frac{1}{2}} - P_{+\frac{1}{2}}^2}.
\end{equation}
In  Figs.~\ref{fig:exp_fastmod}-\ref{fig:exp_slow_resonant}    we
display the experimentally measured
transverse magnetization data represented by $P_{\perp}$ and compare them to
theory by transforming the predictions according to
Eq. (\ref{experiment}).
The initial peak of the non-modulated Rabi-oscillation data, which appears in black in each figure, is used to define $P_{\perp} = 1$ and to normalize the corresponding modulated data.

We begin with the fast-modulation regime, $\omega_{\s m} \! \gg \!
\Omega_{\s R}$, studied in Sect.~\ref{Fast Modulation},
The picture in the ``rotating frame" that results
from the application of the RWA, is that of an average nutation of
the magnetization in the $yz$-plane, superimposed with much faster
wiggles, which for small modulation amplitudes are transverse to the
$yz$-plane. For larger modulation amplitudes, the wiggles move the
magnetization appreciably along the surface of the Bloch sphere,
giving rise to a time-average decrease in the component of the
magnetization that is in the $yz$-plane and subject to a torque
generated by $B_{\s 1}$. This leads directly to the slowing down of the
Rabi nutation expressed by Eq.~\eqref{Modiefied}\cite{Book1}. The wiggles also lead to the fast
modulation component at $\omega_{\s m}$, expressed in Eq.~\eqref{correction}, that is
superimposed on the slowed-down Rabi oscillations. Fig.~\ref{fig:exp_fastmod}a
demonstrates that the corresponding experiments are sensitive to
even the small decrease in the effective Rabi frequency $\Omega_{\s R}
J_{\s 0}\left(\frac{\varepsilon_{\s m}}{\omega_{\s m}}\right)$ that appears for
weak modulation. We were able to fit this data with Eq.~\eqref{Modiefied}
multiplied by a
a decaying exponential that
accounts for the $T_{\s 2}^*$ decay.
The Rabi frequency $\Omega_{\s R}$ and modulation amplitude $\varepsilon_{\s m}$ were experimentally determined, respectively, by examining the non-modulated data and by measuring the current in the modulation coils.

Fig.~\ref{fig:exp_fastmod}b shows data in the regime of both fast
and strong modulation $\varepsilon_{\s m},\, \omega_{\s m} \! \gg \! \Omega_{\s R}$,
where corrections found in Eq.~\eqref{correction} become apparent.
As predicted, the
slowing-down effect on the Rabi oscillations now becomes pronounced,
and the small modulation at frequency $\omega_{\s m}$ described by Eq.~\eqref{correction}
rides on top of the envelope given by Eq.~\eqref{Modiefied}. A peculiar effect seen
in the data and predicted in the theory is the leveling-off of the
magnetization near the nutation angle $\pi/2$, where the
magnetization remains pinned on the
$xy$-plane. In the fit
of the data to
Eq. \eqref{correction}
only $\varepsilon_{\s m}$ was used as a free parameter; for various technical reasons,  larger modulation amplitudes were more difficult to measure experimentally.
The Rabi frequency $\Omega_{\s R}$ as determined from the unmodulated
data was a fixed parameter in the fit.
We note that the $\approx$25\% reduction (due to smaller $B_{\s 1}$) in the value of $\Omega_{\s R}$ determined from the non-modulated data appears to have eliminated
the effects of $T_{\s 2}^*$ decay in the unmodulated
data of Fig.~\ref{fig:exp_fastmod}b as compared to similar data in Fig.~\ref{fig:exp_fastmod}a.


In the strong-modulation regime $\varepsilon_{\s m} \!\gg \! \Omega_{\s R} \!
\gg \! \omega_{\s m}$ studied in Sect.~\ref{Strong Modulation}, the
picture in the rotating frame is that the slowly sweeping modulation
field brings the spins into resonance for only a short fraction of
the modulation period; in the remaining time the modulation field is
so large that the effective field in the rotating frame lies along
the $z$-axis---the spins are essentially out of resonance, and do not
nutate.  Our experiments had $\varepsilon_{\s m}$ at most only a few
times $\Omega_{\s R}$ and thus did not achieve the limit $\varepsilon_{\s m}
\!\gg \! \Omega_{\s R}$ to sufficient degree to turn off the nutation
completely, even when $|\varepsilon(t)|$ was near a maximum. However,
the seemingly complicated data seen in the lower half of Fig.~\ref{fig:exp_slowstrong}
 can still be
understood to a significant degree. Non-trivial behavior of the Rabi
oscillations occurs when $\varepsilon(t)$ is near zero with
periodicity $\pi/\omega_{\s m} \approx 0.166$~ms, as per Eq.~\eqref{tk}.
%
These critical regions of time are marked by dramatic changes in the oscillation behavior lasting about one Rabi period $2\pi/\Omega_{\s R}$, whereas there is a relatively
steady-state oscillation for the remainder of the modulation period. In general, the prediction given by Eq. \eqref{probability} in Sect.~\ref{Strong Modulation} can only be applied when the initial spin state is known, which is only true in our case for the critical region
just after $t=0$.

Hence, the upper half of
Fig.~\ref{fig:exp_slowstrong} shows an expanded view of the first $0.05$~ms of this data set,
which was fit to Eq.~\eqref{probability}.
The fit shown uses the experimentally
determined $\Omega_{\s R} = 17.2 \pm 0.1$~kHz and determines
$\varepsilon_{\s m} = 7.0$~G as a free parameter.
Using the values of $\Omega_{\s R}$, $\omega_{\s m}$, and $\varepsilon_{\s m}$,
the parameter $|\nu| = 0.9$ was calculated from Eq.~\eqref{nu}. Using this value
the saturation level shown with dashed  horizontal line was calculated from
Eq.~\eqref{probability}.
We note with respect to the fit that $\varepsilon_{\s m}$ and the other parameters derived from it are extremely sensitive to whether the magnetization reaches the $xy$-plane ($P_{\perp} = 1$). The precision quoted for $\varepsilon(t)$ may thus be somewhat underestimated. Nonetheless, the data and the fit clearly show the decreasing oscillation period characteristic of the parabolic-cylinder functions on which the theory in this regime is based.

The qualitative features of the data in Fig.~\ref{fig:exp_slowstrong} away from the
critical regions can also be understood. As the critical time period near a zero-crossing of $\varepsilon_{\s m}$ ends, the magnetization vector finds itself at
some particular angle to the $z$-axis that would then essentially not change in the limit
$\varepsilon_{\s m} \!\gg \! \Omega_{\s R}$ until the next zero-crossing. Even when this limit is not well satisfied, if the magnetization is close to being in the $xy$-plane, nutation about an effective field that is mostly along the $z$-direction will produce only small second-order changes in the FID amplitude as the spins nutate: a clear example occurs from $0.2$~ms to $0.3$~ms; here, the FID amplitude is very close to maximum and there is a barely perceptible nutation. Larger oscillations occur for lower values of the FID amplitude when the magnetization is well away from the $xy$-plane,
with the largest occurring between about $0.85$~ms and $0.95$~ms. We
note that the dephasing that occurs due to $B_{\s 1}$ inhomogeneity
appears to be decreased by the strong-slow modulation. The
modulation field has the effect of continuously refocusing the spins
in the rotating-frame $yz$-plane, analogous to the way a Hahn echo
refocuses dephased transverse magnetization precessing about an inhomogeneous $B_{\s 0}$.

In our study of the weak-resonant modulation regime $\Omega_{\s R} \approx \omega_{\s m} \!\gg
\varepsilon_{\s m}$ we focus
on the limit  where the $B_{\s 1}$-pulse is on-resonance with the
Zeeman splitting, and the theoretical prediction for the spin dynamics is
given by  Eq.~\eqref{zerodelta}.
In this regime we observe beats in the Rabi oscillations; the depth of the modulation is determined by the parameter $\kappa$. In Fig.~\ref{fig:exp_slow_resonant} we show a typical example; a fit of the modulated data to Eq.~\eqref{zerodelta} is qualitatively reasonable and clearly exhibits the beat envelope. Quantitatively, a high-quality fit over the entire time interval is made difficult by the extreme sensitivity of $\kappa$, which contains
contains a small difference,
$(\omega_{\s m} -\Omega_{\s R}) \ll \Omega_{\s R}$, to experimental errors in $\omega_{\s m}$.
Because of the above complications, we allowed most of the experimental parameters to become floating fit parameters and performed separate fits for
two time intervals: ``early" and ``late", as shown in
Fig. \ref{fig:exp_slow_resonant}.
The values of $\kappa$ extracted from both fits were
reasonably close to each other: $0.54\pm0.14$ for early and $0.40
\pm 0.12$ for late fits. We attribute the discrepancy
to the slow drift with time of the
Rabi frequency and the modulation-pulse parameters.
The evidence that this drift affects the fit
can already be inferred from Fig.~\ref{fig:exp_fastmod} for the fast-modulation regime.
We note that, as was true in the strong-slow modulation regime, dephasing due to $B_{\s 1}$ inhomogeneity is suppressed by the modulation field.

\section{Concluding remarks}

The main result of the present paper is the mapping Eq. (\ref{correspondence})
which allows one to establish how a spin-$\frac{1}{2}$ system under a
weak resonant ac drive responds to a weak modulation of the longitudinal field
with magnitude $\varepsilon_{\s m}$ and frequency $\omega_{\s m}$ both smaller
than the Zeeman splitting $\Delta_{\s Z}$. This response is strong and leads to
a dramatic modification of the Rabi oscillations  when $\varepsilon_{\s m}$,
remaining much smaller than $\Delta_{\s Z}$, exceeds the Rabi frequency,
$\Omega_{\s R}$. 
Another instance when the response to modulation is strong corresponds to
$\varepsilon_{\s m} \ll \Omega_{\s R}$ but $\omega_{\s m}$ is close to
$\Omega_{\s R}$. Then, within a narrow band,
$|\omega_{\s m}-\Omega_{\s R}|\sim \varepsilon_{\s m}$,
the Rabi oscillations develop an envelope  whose shape
is very sensitive to the detuning of the  $B_{\s 1}$ driving frequency from $\Delta_{\s Z}$.
Experimentally, we have verified qualitatively and quantitatively these non-trivial modifications to Rabi nutation across a broad range of modulation-field frequency and amplitude with fairly simple NMR experiments on protons in water.

When both $\varepsilon_{\s m}$ and $\omega_{\s m}$ are smaller
than $\Omega_{\s R}$, the Rabi oscillations are, in general,
unaffected by the modulation, but even in this domain the
oscillations acquire a fully developed envelope in the vicinities of
 certain modulation frequencies  $\omega_{\s m}^{(p)}=\Omega_{\s R}/(2p+1)$.
With increasing $p$ the period of the envelope increases, while the
domain, $\omega_{\s m}^{(p)}-\omega_{\s m}$, where it develops
progressively shrinks, see Eq. (\ref{fromshirley}). Physically, the limit
$p\gg 1$ corresponds to a  multiphoton process where the role of
photons is played by the ``quanta of modulation". This limit is
easily captured within the adiabatic approximation\cite{Ostrovsky}.

In certain semiconductor materials of interest, where magnetic resonance is detected electrically with the help
of the
pulsed technique\cite{BoehmeLips}, the effect of modulation can be
even more peculiar. This is because the change
of conductivity, measured in experiment, is determined
by synchronized Rabi oscillations in {\em pairs} of spins,
and reflects the beating of these oscillations between the
pair components\cite{BoehmeOrganic}.

As a final remark, note that there is a conceptual similarity between the second-order Rabi oscillations,
which develop at  $\omega_{{\scriptscriptstyle m}}=\Omega_{{\scriptscriptstyle R}}$ and
 Rabi-vibronic resonance\cite{Rachel}.
In the latter case, the  modulation of spacing between the Zeeman levels is accomplished via
coupling of two-level system to a harmonic oscillator with frequency
close to  $\Omega_{{\scriptscriptstyle R}}$.

\begin{acknowledgements}
We acknowledge useful discussions with V.~V. Mkhitaryan and G.~Laicher.
 We acknowledge the support of this work by the National Science Foundation through the Materials Research Science and Excellence Center
(\#DMR-1121252).
\end{acknowledgements}

\end{document}